\documentclass[twocolumn,prd,showpacs,preprintnumbers,amsmath,amssymb,floatfix]{revtex4}

\usepackage{graphicx,epsf}

\usepackage{bm}
\usepackage{amsfonts}
\usepackage{lineno,hyperref}
\usepackage{float,natbib}
\usepackage{epstopdf,color}

\newcommand*\DAlambert{\mathop{}\!\mathbin\Box}

\begin{document}

\title{Anisotropic strange stars under simplest minimal matter-geometry coupling in the $f\left(R,\mathcal{T}\right)$ gravity}

\author{Debabrata Deb}
\email{ddeb.rs2016@physics.iiests.ac.in}
\affiliation{Department of Physics, Indian Institute of Engineering Science 
	and Technology, Shibpur, Howrah 711103, West Bengal, India}

\author{Farook Rahaman}
\email{rahaman@associates.iucaa.in}
\affiliation{Department of Mathematics, Jadavpur University, Kolkata 700032, West Bengal, India}

\author{Saibal Ray}
\email{saibal@associates.iucaa.in}
\affiliation{Department of Physics, Government College of Engineering and
Ceramic Technology, Kolkata 700010, West Bengal, India}

\author{B.K. Guha}
 \email{bkguhaphys@gmail.com}
\affiliation{Department of Physics, Indian Institute of Engineering Science 
	and Technology, Shibpur, Howrah 711103, West Bengal, India}

\date{\today}

\begin{abstract}
We study strange stars in the framework of $f\left(R,\mathcal{T}\right)$ theory of gravity. To provide exact solutions of the field equations it is considered that the gravitational Lagrangian can be expressed as the linear function of the Ricci scalar $R$ and the trace of the stress-energy tensor $\mathcal{T}$, i.e. $f(R,\mathcal{T})=R+2\chi\mathcal{T}$, where $\chi$ is a constant. We also consider that the strange quark matter (SQM) distribution inside the stellar system is governed by the phenomenological MIT Bag model equation of state (EOS), given as $p_r=\frac{1}{3}\left(\rho-4\,B \right)$, where $B$ is the Bag constant. Further, for a specific value of $B$ and observed values of mass of the strange star candidates we obtain the exact solution of the modified Tolman-Oppenheimer-Volkoff (TOV) equation in the framework of $f\left(R,\mathcal{T}\right)$ gravity and have studied in detail the dependence of the different physical parameters, like the metric potentials, energy density, radial and tangential pressures and anisotropy etc., due to the chosen different values of $\chi$. Likewise in GR, as have been shown in our previous work~[Deb et al., Ann. Phys. (Amsterdam) 387, 239 (2017)] in the present work also we find maximum anisotropy at the surface which seems an inherent property of the strange stars in modified $f\left(R, \mathcal{T}\right)$ theory of gravity. To check the physical acceptability and stability of the stellar system based on the obtained solutions we have performed different physical tests, viz., the energy conditions, Herrera cracking concept, adiabatic index etc. In this work, we also have explained the effects, those are arising due to the interaction between the matter and the curvature terms in $f\left(R,\mathcal{T}\right)$ gravity, on the anisotropic compact stellar system. It is interesting to note that as the values of $\chi$ increase the strange stars become more massive and their radius increase gradually so that eventually they gradually turn into less dense compact objects. The present study reveals that the modified $f\left(R,\mathcal{T}\right)$ gravity is a suitable theory to explain massive stellar systems like recent magnetars, massive pulsars and super-Chandrasekhar stars, which can not be explained in the framework of GR. However, for $\chi=0$ the standard results of Einsteinian gravity are retrieved.
\end{abstract}

\pacs{ 04.20.Jb, 95.30.Sf, 04.50.Kd, 04.40.Dg}

\maketitle

\section{Introduction}
We are living in the age of the accelerated expansion of the universe which is well supported by the evidences of recent observations, like CMB, LSS, supernovae-Ia and BAO~\cite{Riess1998,Perlmutter1999,Bernardis2000,Perlmutter2003}. Thus the modern cosmology is mainly dependent on the recent observational evidences of the accelerated expansion of the universe. However, Einstein's general theory of relativity failed to answer the satisfactory reason behind this accelerated expansion of the universe. Although in this connection many researchers~\cite{Caldwell2002,Nojiri2003,Odinstov2003,Padmanabhan2002,Kamenshchik2001,Bento2002} predicted that the sole reason behind this phenomenon is the presence of an unknown form of exotic energy dominated by the negative pressure which is widely known as the dark energy. The gravitational interaction is the most fundamental but least understood force of the nature. According to the strings/M-theory (also known as theory of everything) general relativity is an approximation and consistent to the small curvature. Though in the early days some unknown gravitational theory described the evolution of the universe but now it is well accepted that the modified gravity which is a classical generalization of the general relativity, can explain the early-time inflation and the late-time acceleration without introducing any form of the dark component. Also, some of the modified gravity theories with the gravitational term are well valid in the high energy realm which produced inflationary epoch. The curvature decreases during the evolution of the universe and in the intermediate universe general relativity provides a sufficient approximation. Interestingly, the early-time as well as the late-time acceleration happen due to the fact that some sub-dominant terms of gravitational action may become essential to the large or small curvatures. Though the complete gravitational action should be described by some fundamental theory which is yet to be achieved, but such approach of the alternative theory of gravity can be considered as a dynamical solution of the cosmological constant problem. The modified gravity approach in the absence of the fundamental quantum gravity showed a promising way out as it is well consistent with the observational data and data from local tests~\cite{Nojiri2011}. Few well-known relevant alternative gravity theories are $f \left(R\right)$ gravity~\cite{Nojiri2003a,Carroll2004,Allemandi2005,Nojiri2007,Bertolami2007}, Brans-Dicke~(BD), $f\left(G\right)$~\cite{Bamba2010a,Bamba2010b,Rodrigues2014} gravity, $f\left(\mathbb{T}\right)$ gravity~\cite{Bengocheu2009,Linder2010}, scalar tensor theories of gravity and $f\left(R,G\right)$ gravity, etc., where $R$, $G$ and $\mathbb{T}$ are the scalar curvature, the Gauss-Bonnet scalar and the torsion scalar, respectively. 

In his pioneering work Capozziello~\cite{Capozziello2002} proposed a new modified theory of gravity to tackle the issue of dark energy.  Later, Allemandi et al.~\cite{Allemandi2005} have introduced the nonlinear scalar-gravity theories in the Palatini formulation. In their important review article Nojiri and Odintsov~\cite{Nojiri2011} have presented a detailed study on the various extended gravity models, viz., traditional $f\left(R\right)$ and Ho{\v r}ava-Lifshitz $f\left(R\right)$ gravity, scalar-tensor theory, string-inspired and Gauss–Bonnet theory, non-local gravity, non-minimally coupled models, and power-counting renormalizable covariant gravity. In this large volume of works they have investigated relation between the discussed modified gravity theories and their different representations. Further, the authors also have demonstrated how these extended gravity theories are showing well agreement with the local tests and featuring well justified description of the inflation with the dark energy epoch. Again, Capozziello and Laurentis~\cite{Capozziello2011} presented an extended study on the different modified theories of gravity, viz., $f\left(R\right)$ gravity, scalar-tensor gravity, Brans-Dicke gravity and $f\left(R,\phi\right)$ gravity, etc., to address the shortcomings of GR at the scale of ultraviolet and infrared. Astashenok et al.~\cite{Astashenok2013} and Capozziello~\cite{Capozziello2016} presented models for neutron stars under different form of the $f\left(R\right)$ gravity. In another work Astashenok et al.~\cite{Astashenok2015} have studied non-perturbative models for strange quark stars in $f\left(R\right)$ gravity.

Recently Harko et al.~\cite{Harko2011} presented a more generalized form of $f \left(R\right)$ gravity theory by choosing the matter Lagrangian consists of an arbitrary function of the Ricci scalar ($R$) and the trace of the energy momentum tensor ($\mathcal{T}$) given as $f\left(R,\mathcal{T}\right)$. This is known as $f\left(R,\mathcal{T}\right)$ theory of gravity. Immediately it has drawn attention of many researchers and in the framework of many cosmological models~\cite{Myrzakulov2012,Jamil2012,Shabani2013,Shabani2014,Moraes2015d,Momeni2015,Zaregonbadi2016,Shabani2017a,Shabani2017b} have been studied. Besides cosmology this gravity has successfully been studied in the realm of astrophysics too.  Under astrophysics it is observed that Sharif et al.~\cite{sharif2014} explored the factors that affect the stability of a locally isotropic spherically symmetric self gravitating system. By employing the perturbation scheme Noureen et al.~\cite{noureen2015,noureen2015b,noureen2015c} have presented a series of works on the dynamical instability of spherically symmetric anisotropic collapsing stars under different conditions. Further, Zubair and Noureen~\cite{zubair2015a} studied the dynamical stability of axially symmetric anisotropic sources whereas Zubair et al.~\cite{zubair2015b} investigated the possible formation of compact stars by employing Krori and Barua metric. Alhamzawi and Alhamzawi~\cite{Ahmed2015} have shown the effect of $f\left(R,\mathcal{T}\right)$ gravity on the gravitational lensing and also compared their result with the standard results of general relativity (GR). 

Furthermore, general relativity and its possible extension~\cite{Psaltis2008} can be distinguished due to the strong gravitational field regimes of the relativistic stars. Various developments of the new stellar structures constitute the signature of the extended gravity model~\cite{Capozziello2011a,Capozziello2012} as they have important observational consequences. Also, in particular some simplest extension of the general relativity, for example $f\left(R\right)$ gravity do not support existence of the stable stellar system~\cite{Briscese2007,Abdalla2005,Bamba2008,Kobayashi2008,Babichev2010,Nojiri2009,Bamba2011}. On the other hand, the stability of the stellar system in modified gravity in the certain cases can be achieved using the so-called Chameleon Mechanism~\cite{Khoury2004a,Khoury2004b,Upadhye2009}.

Although all the above-mentioned literature are studied on the basis of the analytical solution Moraes et al.~\cite{Moraes2015} first presented the exact solution of the Tolman-Oppenheimer-Volkoff (TOV) equation in $f(R,\mathcal{T})$ gravity, using Runge-Kutta 4th-order method and studied hydrostatic equilibrium configurations for neutron stars and strange stars. Here, we would like to mention that unfortunately in the TOV equation~[Eq. $\left(3.9\right)$] a minus sign has been missed in their paper~\cite{Moraes2015}. Using the results of Moraes et al.~\cite{Moraes2015} later on Das et al.~\cite{Amit2016} presented an analytical model of compact stars in $f\left(R,\mathcal{T}\right)$ gravity by employing the Lie algebra with the conformal Killing vectors. However, in another work on gravastars Das et al.~\cite{Amit2017} have corrected the form of the TOV equation and provided an analytical model in $f\left(R,\mathcal{T}\right)$ gravity.

Harko et al.~\cite{Harko2011} in their pioneering work mentioned that the motivation behind considering $T$-dependence in the $f\left(R,\mathcal{T}\right)$ theory of gravity is the possible existence of exotic imperfect fluids or quantum effects, such as the particle production~\cite{Harko2014}. The authors in their study~\cite{Harko2011} showed that the covariant derivative of the energy-momentum tensor is not zero and an extra acceleration will always be present in $f\left(R,\mathcal{T}\right)$ gravity due to the coupling between the matter and the curvature terms. Hence particles will follow non-geodesic path in $f\left(R,\mathcal{T}\right)$ gravity. Later, Chakraborty~\cite{SC2013} addressed this issue and showed that for a specific form of the function $f\left(R,\mathcal{T}\right)$, as $f\left(R,\mathcal{T}\right)=R+h\left(\mathcal{T}\right)$, the test particles follow the geodesic path. Hence the author~\cite{SC2013} demonstrated that the whole system would act like non-interacting two fluid system where the second type of fluid is originated due to the interaction between the geometry and the matter.

Here we would like to highlight the fact that in the framework of GR one can find a vast number of works~\cite{Ivanov2002,SM2003,MH2003,Usov2004,Varela2010,Rahaman2010,Rahaman2011,Rahaman2012,Kalam2012,Deb2016,Shee2016,Maurya2016,Maurya2017}, where influence of the anisotropy on the static spherically symmetric compact objects have been studied. It is to note that when the radial component of the pressure, ${p_r}(r)$, differs from the angular component, ${p_{\theta}}(r) = {p_{\phi}}(r) \equiv {p_{t}}(r)$ the system can be said anisotropic in nature. Clearly, the condition ${p_{\theta}}(r) = {p_{\phi}}(r)$ is rising due to the effect of the spherical symmetry. In a physical system, the pressures are anisotropic when the associated scalar field has a non-zero spatial gradient. The anisotropic stress in the present case may be arising due to the presence of the anisotropic nature of the two-fluid system.

We have arranged the present article as follows: Basic mathematical formulation of $f\left(R,\mathcal{T}\right)$ gravity is presented in Sec.~\ref{sec1}. In Sec.~\ref{sec2} we formulate basic stellar equations and present the solution of the Einstein field equations in Sec.~\ref{sec3}. We examine physical acceptability and stability of the stellar system in Sec.~\ref{sec4} by studying energy conditions~\ref{subsec4.1}, mass-radius relation~\ref{subsec4.2},~stability of the stellar model~\ref{subsec4.3} and compactification factor as well as reddshift~\ref{subsec4.4}. Finally, we conclude our study with a discussion in Sec.~\ref{sec5}.

\section{Basic formulation of $f\left(R,\mathcal{T}\right)$ theory of gravity}\label{sec1}

Following Harko et al.~\cite{Harko2011}, the modified form of Einstein-Hilbert (EH) action in $f\left(R,\mathcal{T}\right)$ gravity reads
\begin{equation}\label{1.1}
S=\frac{1}{16\pi}\int d^{4}xf(R,\mathcal{T})\sqrt{-g}+\int
d^{4}x\mathcal{L}_m\sqrt{-g},
\end{equation}
where $g$ and $\mathcal{L}_m$ are the determinant of the metric $g_{\mu\nu}$ and the matter Lagrangian density, respectively. We adopt throughout the article $G=1=c$.

Variation of the modified EH action~(\ref{1.1}) in $f\left(R,\mathcal{T}\right)$ gravity with respect to $g_{\mu\nu}$ yields the modified field equation as follows
\begin{eqnarray}\label{1.2}
&\qquad\hspace{-3cm} G_{\mu\nu}=\frac{1}{{f_R}\left(R,\mathcal{T}\right)}\Big[\lbrace{8\pi+f_\mathcal{T}}\left(R,\mathcal{T}\right)\rbrace{T_{\mu\nu}}\nonumber \\
&\qquad -\rho{g_{\mu\nu}}{f_\mathcal{T}}\left(R,\mathcal{T}\right)+\frac{1}{2}\lbrace{f_\mathcal{T}}\left(R,\mathcal{T}\right)-R{f_R}\left(R,\mathcal{T}\right)\rbrace{g_{\mu\nu}}\nonumber\\ 
&\qquad\hspace{1cm} +\left(\nabla_{\mu} \nabla_{\nu}-g_{\mu\nu}\DAlambert\right) {{f_R}\left(R,\mathcal{T}\right)}\Big],
\end{eqnarray}
where $f_R (R,\mathcal{T})=\partial f(R,\mathcal{T})/\partial R$~and~$f_\mathcal{T}(R,\mathcal{T})=\partial f(R,\mathcal{T})/\partial\mathcal{T}$ whereas $\DAlambert \equiv\partial_{\mu}(\sqrt{-g} g^{\mu\nu} \partial_{\nu})/\sqrt{-g}$ is the D'Alambert operator and $R_{\mu\nu}$ is the Ricci tensor. We assume that $\mathcal{L}_m=\rho$, where $\rho$ is the energy density of SQM distribution. 

Now we define $T_{\mu\nu}$, which represents the stress-energy tensor for the anisotropic fluid distribution, in the following form
\begin{eqnarray}\label{1.3}
T_{\mu\nu}=\left(\rho+{p_t}\right){u_{\mu}}{u_{\nu}}-{p_t}{g_{\mu\nu}}+\left({p_r}-{p_t}\right){v_{\mu}}{v_{\nu}},
\end{eqnarray}
where $p_r$ and $p_t$ represent the radial and tangential pressures of the SQM distribution, respectively whereas $u_{\mu}$ and $v_{\mu}$ represent four-velocity and radial four-vector, respectively.

The covariant divergence of the stress-energy tensor (\ref{1.3}) is given by
\begin{eqnarray}\label{1.4}
&\qquad\hspace{-0.5cm}\nabla^{\mu}T_{\mu\nu}=\frac{f_\mathcal{T}(R,\mathcal{T})}{8\pi -f_\mathcal{T}(R,\mathcal{T})}\big[\left(-T_{\mu\nu}+\rho{g_{\mu\nu}}\right)\nabla^{\mu}\ln f_\mathcal{T}(R,\mathcal{T}) \nonumber\\
&\qquad\hspace{-0.5cm}-2\nabla^{\mu}T_{\mu\nu}+\frac{1}{2}{g_{\mu\nu}}{\nabla^{\mu}}\left(2\rho-\mathcal{T}\right)\big].
\end{eqnarray}

Following Harko et al.~\cite{Harko2011}, in the present article we consider simple linear form of the function $f\left(R,\mathcal{T}\right)$ as $f\left(R,\mathcal{T}\right)=f\left(R\right)+2f\left(\mathcal{T}\right)$, where $f\left(R\right)=R$ and $f\left(\mathcal{T}\right)=2\chi \mathcal{T}$. This form of the function $f\left(R,\mathcal{T}\right)$ has been broadly used by several authors \cite{singh2015,moraes2014b,moraes2015a,moraes2015b,moraes2017,singh2014,baffou2015,shabani2013,shabani2014,sharif2014b,reddy2013b,kumar2015,shamir2015,Fayaz2016}.

Now substituting the assumed form of the function $f\left(R,\mathcal{T}\right)$ into Eq.~(\ref{1.2}) we find
\begin{eqnarray}\label{1.5}
G_{\mu\nu}=8\pi T_{\mu\nu}+\chi \mathcal{T}g_{\mu\nu}+2\chi(T_{\mu\nu}-\rho g_{\mu\nu})=8\pi T_{\mu\nu,eff},\nonumber\\
\end{eqnarray}
where $G_{\mu\nu}$ is the standard Einstein tensor and $T_{\mu\nu,eff}=T_{\mu\nu}+\frac{\chi}{8\pi} \mathcal{T}g_{\mu\nu}+\frac{\chi}{4\pi}(T_{\mu\nu}-\rho g_{\mu\nu})$. The usual general relativistic results can be achieved by substituting $\chi=0$ into Eq.~(\ref{1.5}). 

Now, substituting $f(R,\mathcal{T})=R+2\chi\mathcal{T}$ in Eq.~(\ref{1.4}) we have
\begin{eqnarray}\label{1.6}
\left(4\pi+\chi\right)\nabla^{\mu}T_{\mu\nu}=-\frac{1}{2}\chi\left[g_{\mu\nu}\nabla^{\mu}\mathcal{T}-2\,\nabla^{\mu}(\rho g_{\mu\nu})\right].
\end{eqnarray}

We can write Eq.~(\ref{1.6}) as follows
\begin{equation}\label{1.7}
\nabla^{\mu}T_{\mu\nu,eff}=0.
\end{equation}

Here also one may achieve the standard form of the conservation of stress-energy tensor as GR by substituting $\chi=0$ into Eq.~(\ref{1.6}).

\section{Basic stellar equations in $\bf{\textit{f}\left(R,\mathcal{T}\right)}$ theory of gravity}\label{sec2}
We consider the spherically symmetric metric in its usual form 
\begin{equation}\label{2.1}
ds^2=e^{\nu(r)}dt^2-e^{\lambda(r)}dr^2-r^2(d\theta^2+\sin^2\theta
d\phi^2),
\end{equation}
 where $\nu$ and $\lambda$ are metric potentials and function of the radial coordinate only.

Hence, using Eqs.~(\ref{1.3}),~(\ref{1.5})~and~(\ref{2.1}) we find the Einstein field equations for the spherically symmetric anisotropic stellar system given as
\begin{eqnarray}\label{2.2}
&\qquad\hspace{-2cm} {{\rm e}^{-\lambda}} \left( {\frac {\lambda^{{\prime}}}{r}}-\frac{1}{{r}^{2}} \right) +\frac{1}{{r}^{2}}=\left( 8\,\pi +\chi \right) \rho-\chi\,p_{{r}}-2\,\chi\,p_{{t}}\nonumber\\
&\qquad\hspace{-1.7cm} =8\pi{{\rho}_{eff}}, \\ \label{2.3}
&\qquad\hspace{-2cm} {{\rm e}^{-\lambda}} \left( {\frac {\nu^{{\prime}}}{r}}+\frac{1}{{r}^{2}} \right) -\frac{1}{{r}^{2}}=\chi\,\rho+ \left( 8\,\pi +3\,\chi \right) p_{{r}}+2\,\chi\,p_{{t}}\nonumber\\
&\qquad\hspace{-1.8cm}=8\pi{{p}_{effr}}, \\ \label{2.4}
&\qquad\hspace{-1.5cm} \frac{{\rm e}^{-\lambda}}{2}\left( \nu^{{\prime\prime}}+\frac{{{\nu^{{\prime}}}}^{2}}{2}+{\frac {\nu^{{\prime}}-\lambda^{{\prime}}}{r}}-\frac{\nu^{{\prime}}\lambda^{{\prime}}}{2} \right) =\chi\,\rho+\chi\,p_{{r}}+ \left( 8\,\pi +4\,\chi \right) p_{{t}}\nonumber\\
&\qquad\hspace{0.5cm}=8\pi{{p}_{efft}},
\end{eqnarray}
where a $\prime$ denotes differentiation with respect to the radial coordinate $r$. Here ${\rho}_{eff}$,~${p}_{effr}$~and~$p_{efft}$ represents the effective energy density, radial pressure and tangential pressure for our system and given as
\begin{eqnarray}\label{2.5}
{{\rho}_{eff}}=\rho+{\frac {\chi }{8\pi }}\left( \rho-p_{{r}}-2\,p_{{t}} \right),\\ \label{2.6}
{{p}_{effr}}=p_{{r}}+{\frac {\chi }{8\pi }}\left( \rho+3\,p_{{r}}+2\,p_{{t}} \right),\\ \label{2.7}
{{p}_{efft}}=p_{{t}}+{\frac {\chi}{8\pi }\left( \rho+p_{{r}}+4\,p_{{t}} \right)}.
\end{eqnarray}

We assume that the SQM distribution inside the strange stars is governed by the simple phenomenological MIT Bag model EOS~\cite{Chodos1974}. In bag model, by introducing {\it ad hoc} bag function all the corrections of energy and pressure functions of SQM have been maintained. We also consider that the quarks are non-interacting and massless in a simplified bag model. The quark pressure therefore can be defined as 
\begin{equation}\label{2.8}
{p_r}={\sum_{f=u,d,s}}{p^f}-{B},
 \end{equation}
where $p^f$ is the individual pressure of the up~$\left(u\right)$, down~$\left(d\right)$ and strange~$\left(s\right)$ quark flavors and $B$ is the vacuum energy density (also well known as Bag constant) which is a constant quantity within a numerical range. In the present article we consider the value of Bag constant as $B=83~MeV/{{fm}^3}$~\cite{Rahaman2014}.

Now the individual quark pressure ($p^f$) can be defined as $p^f=\frac{1}{3}{{\rho}^f}$, where ${{\rho}^f}$ is the energy density of the individual quark flavor. Hence, the energy density, $\rho$ of the de-confined quarks inside the bag is given by
\begin{equation}
{{\rho}}={\sum_{f=u,d,s}}{{\rho}^f}+B. \label{2.9}
\end{equation}

Using Eqs.~(\ref{2.8})~and~(\ref{2.9}) we have the EOS for SQM given as
\begin{equation}
{p_r}=\frac{1}{3}({{\rho}}-4B).\label{2.10}
\end{equation} 

It is observed that ignoring critical aspects of the quantum particle physics in the framework of GR several authors~\cite{1,2,3,4,5,6,7,8} successfully have been introduced this simplified form of the MIT Bag EOS to study stellar systems made of SQM. 

To have non-singular monotonically decreasing matter density inside the spherically symmetric stellar system, following Mak and Harko~\cite{Harko2002}, we assume simplified form of $\rho$ given as
\begin{equation}\label{2.11}
\rho(r)=\rho_c\left[1-\left(1-\frac{\rho_0}{\rho_c}\right)\frac{r^{2}}{R^{2}}\right],
\end{equation}
where $\rho_c$ and $\rho_0$ are constants and denote the maximum and minimum values of $\rho$ at the center and on the surface, respectively.

Now following~\cite{Moraes2017} we consider $p_t$ is related to $\rho$ by a relation given as
\begin{eqnarray}
p_{t}=c_{1}\rho+c_{2},\label{2.11a}
\end{eqnarray}
where $c_1$ and $c_2$ are constants.

We define the mass function of the spherically symmetric stellar system as
\begin{equation}\label{2.12}
m \left( r \right) =4\,\pi\int_{0}^{r}\!{{\rho}_{eff}} \left( r \right) {r}^{2}{dr}.
\end{equation}

At this juncture we consider the Schwarzschild metric to represent the exterior spacetime of our system given as
\begin{eqnarray}\label{2.13}
 {ds}^2=\left(1-\frac{2M}{r}\right)dt^2- \frac{{dr}^2}{\left(1-\frac{2M}{r}\right)}-r^2(d\theta^2+\sin^2\theta
d\phi^2),\nonumber\\
 \end{eqnarray} 
where $M$ is the total mass of the stellar system. 

Now, substituting Eq.~(\ref{2.12}) into Eq.~(\ref{2.2}) we find
\begin{eqnarray}\label{2.14}
 {{\rm e}^{-\lambda \left( r \right) }}=1-{\frac {2m}{r}}.
 \end{eqnarray}

\section{Solution of the Einstein Field equations}\label{sec3}
Solving Einstein field equations (\ref{2.2})-(\ref{2.4}) and using the Eqs.~(\ref{2.5})-(\ref{2.7}),(\ref{2.10})-(\ref{2.12}) we obtain expressions for the different physical parameters, which are given as
\begin{eqnarray}\label{3.1}
&\qquad\hspace{-2cm} \lambda \left(r \right) =-\ln  \left[{\frac {\lambda_{{3}}{r}^
{4}-80 \lambda_{{1}} \left(B\chi+2\pi \rho_{{c}} \right) {r}^{2}
{R}^{2}+60\lambda_{{1}}{R}^{2}}{ 15\left(3\chi c_{{1}}+4\pi +3
\chi \right) {R}^{2}}} \right],\\\label{3.2}
&\qquad\hspace{-1cm} \nu\left(r\right)=\frac {1}{36864 \nu_{{4}} \left(-3 \chi c_{{1}}+12 \pi +\chi \right) } \Bigg[\nu_{{3}}{\rm arctanh} \Big\lbrace\Big(\big[16\nu_{{2}}B{R}^{5}\nonumber\\
&\qquad\hspace{-1cm}-32 B{R}^{3}{r}^{2}\nu_{{2}}+6 M{r}^{2} \left(\pi +\chi \right)  \big] \lambda_{{2}}-5 \lambda_{{1}}M\pi {R}^{2}\Big)/{16\nu_{{4}}}\Big\rbrace\nonumber\\
&\qquad\hspace{-1cm} -294912\big\lbrace\left( \frac{3}{8}c_{{1}}+\frac{1}{2} \right) \chi+\pi  \big\rbrace \nu_{{4}}\ln\Big[384 {R}^{5}{r}^{2}\lambda_{{2}}\nu_{{2}}B\nonumber\\
 &\qquad\hspace{-1cm}-384 \lambda_{{2}}{r}^{4}\nu_{{2}}B{R}^{3}-120{R}^{2}{r}^{2}\lambda_{{1}}M\pi +72 M{r}^{4} \left( \pi +\chi \right) \lambda_{{2}}\nonumber\\
&\qquad\hspace{-1cm}+\nu_{{5}}\Big] -\nu_{{3}}{\rm arctanh}\Big[\big\lbrace -\lambda_{{2}}\nu_{{2}}B{R}^{5}-{\frac {5 \lambda_{{1}}M\pi {R}^{2}}{16}}\nonumber\\
&\qquad\hspace{-1cm}+\frac{3}{8}\lambda_{{2}}{R}^{2}M \left(\pi +\chi \right)\big\rbrace/{\nu_{{4}}}\Big] +442368 \nu_{{4}} \Big\lbrace\Big[ \frac{2}{3}\pi + \left(\frac{1}{4}c_{{1}}+\frac{1}{3}\right) \chi\Big]\nonumber\\
 &\qquad\hspace{-1cm} \ln\Big\lbrace 24 {R}^{4}\nu_{{1}} \left(R-2M \right)  \Big\rbrace +\ln  \left(1-{\frac {2M}{R}} \right) \lambda_{{2}} \Big\rbrace\Bigg], \\\label{3.3}
 &\qquad\hspace{-3.8cm} \rho_{{{\it eff}}}={\frac {12 {{\lambda}_2} \left( \rho_{{0}}-\rho_{{c}} \right) {r}^{2}+6{R}^{2} \left( B\chi+2\pi\rho_{{c}} \right) }{12\pi{R}^{2}}},\\ \label{3.4}
 &\qquad\hspace{-3.5cm} p_{{{\it effr}}}=-{\frac { \left(3\chi c_{{1}}+4 \pi +3\chi \right)  \left( \rho_{{c}}-\rho_{{0}} \right) {r}^{2}+p_{{1}}{R}^{2}}{12\pi{R}^{2}}},\\ \label{3.5}
 &\qquad\hspace{-1cm} p_{{{\it efft}}}= \Big[\Big\lbrace 3\left( c_{{1}}+\frac{1}{3}\right)  \left(\rho_{{0}}-\rho_{{c}} \right) {r}^{2}-5\left( B-\frac{2}{5}\rho_{{c}} \right) {R}^{2}\Big\rbrace \chi\nonumber\\
  &\qquad\hspace{-1cm} -8\pi\Big\lbrace -\frac{3}{4}c_{{1}} \left( \rho_{{0}}-\rho_{{c}} \right) {r}^{2}+{R}^{2} \left(B-\frac{1}{4}\rho_{{c}} \right)\Big\rbrace\Big]\Bigg/6\pi{R}^{2},\nonumber\\
\end{eqnarray}
where $\lambda_{{1}}$,~$\lambda_{{2}}$,~$\lambda_{{3}}$,~$\nu_{{1}}$,~$\nu_{{2}}$,~$\nu_{{3}}$,~$\nu_{{4}}$,~$\nu_{{5}}$,~${\rho}_c$~and~${\rho}_0$ are constants and their expressions are shown in Appendix.


\begin{figure}[!htpb]
\centering
	\includegraphics[width=6cm]{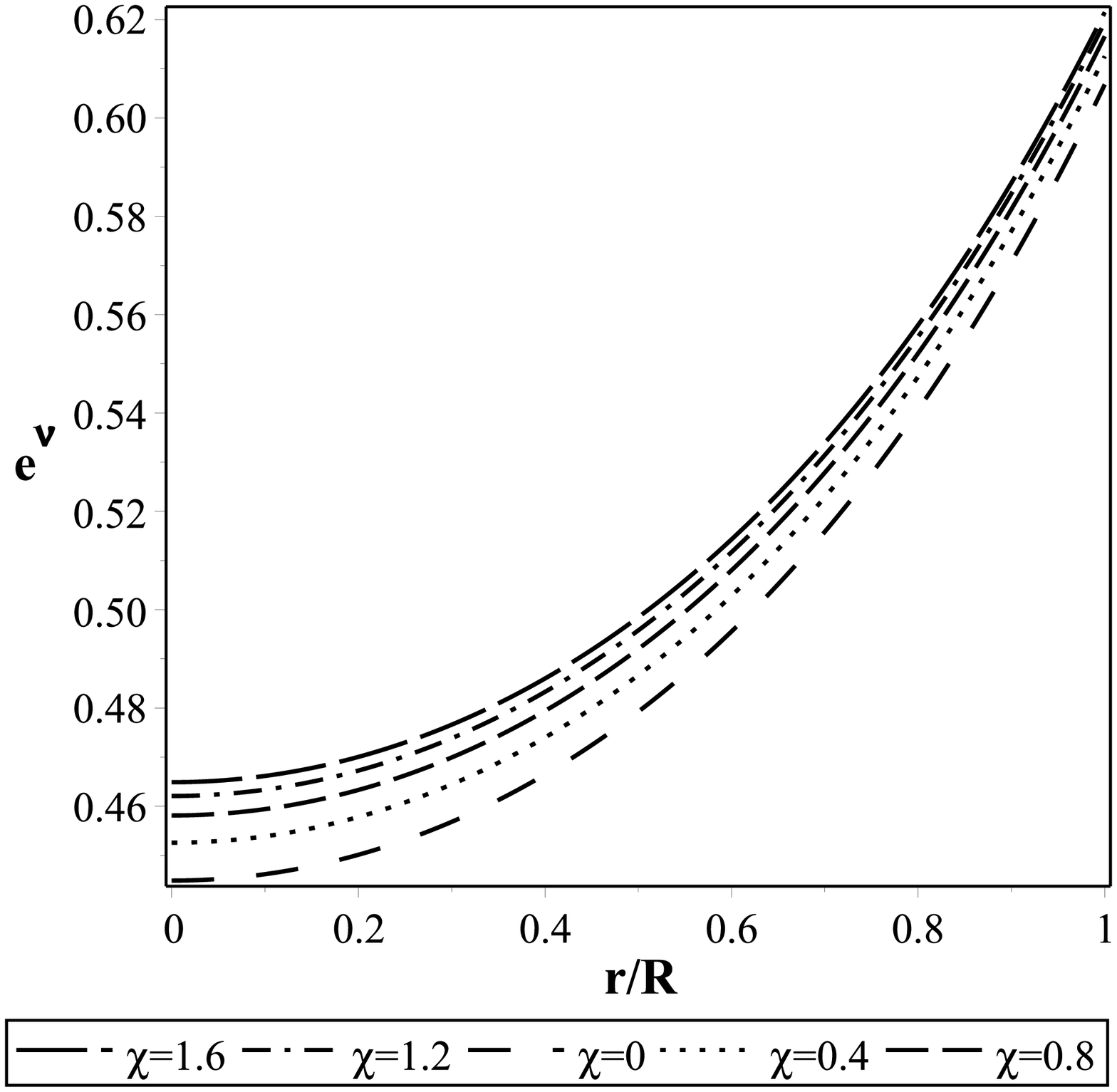}
	\includegraphics[width=6cm]{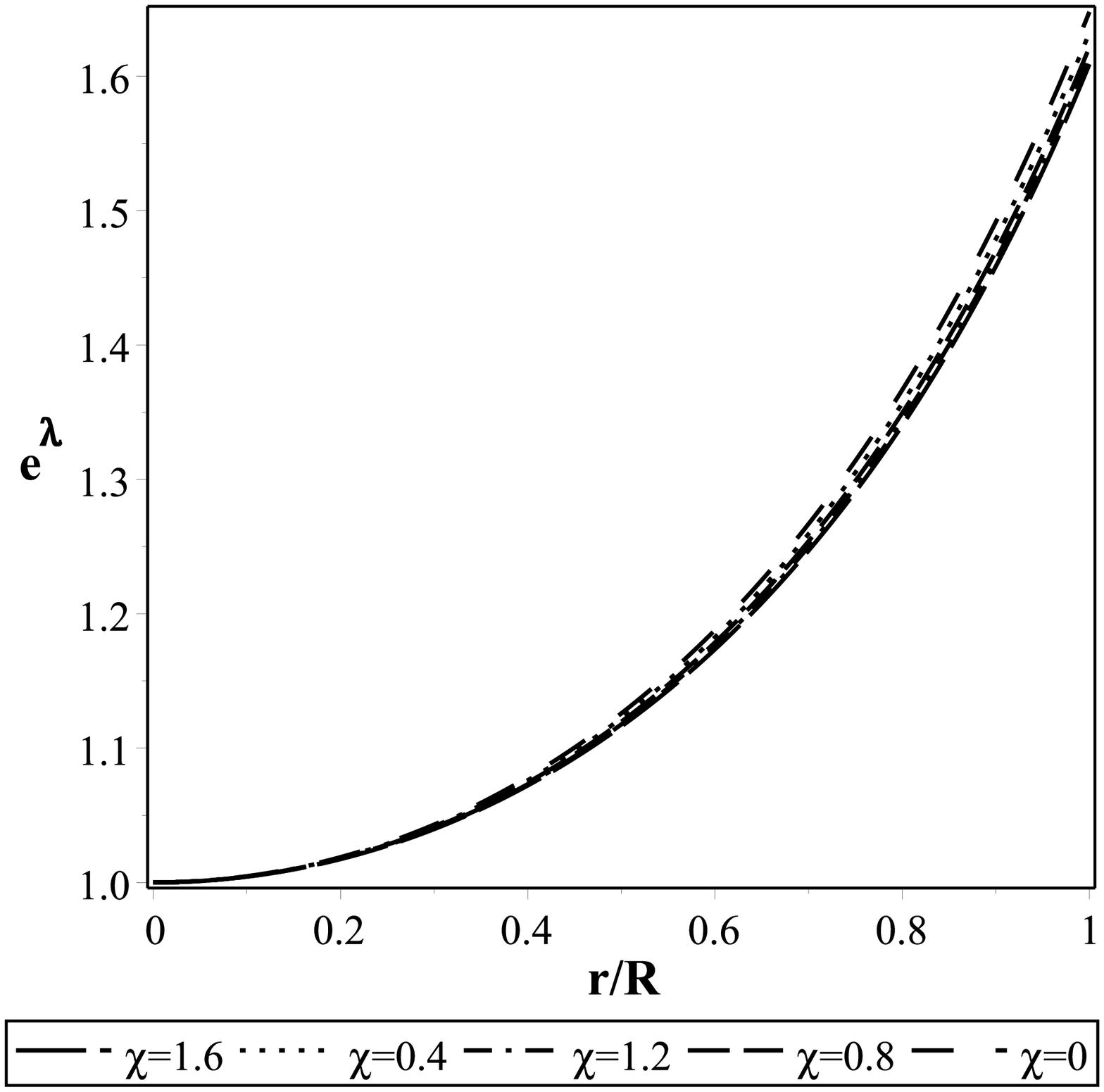}
		\caption{Variation of (i) ${e}^{\nu(r)}$ (upper panel) and (ii) ${e}^{\lambda(r)}$
(lower panel) as a function of the radial coordinate $r/R$ for the strange star $LMC~X-4$. Here $B=83~ MeV/{{fm}^3}$ and $c_1=0.2$.} \label{Fig1}
\end{figure}



\begin{figure}[!htpb]\centering
\includegraphics[width=6cm]{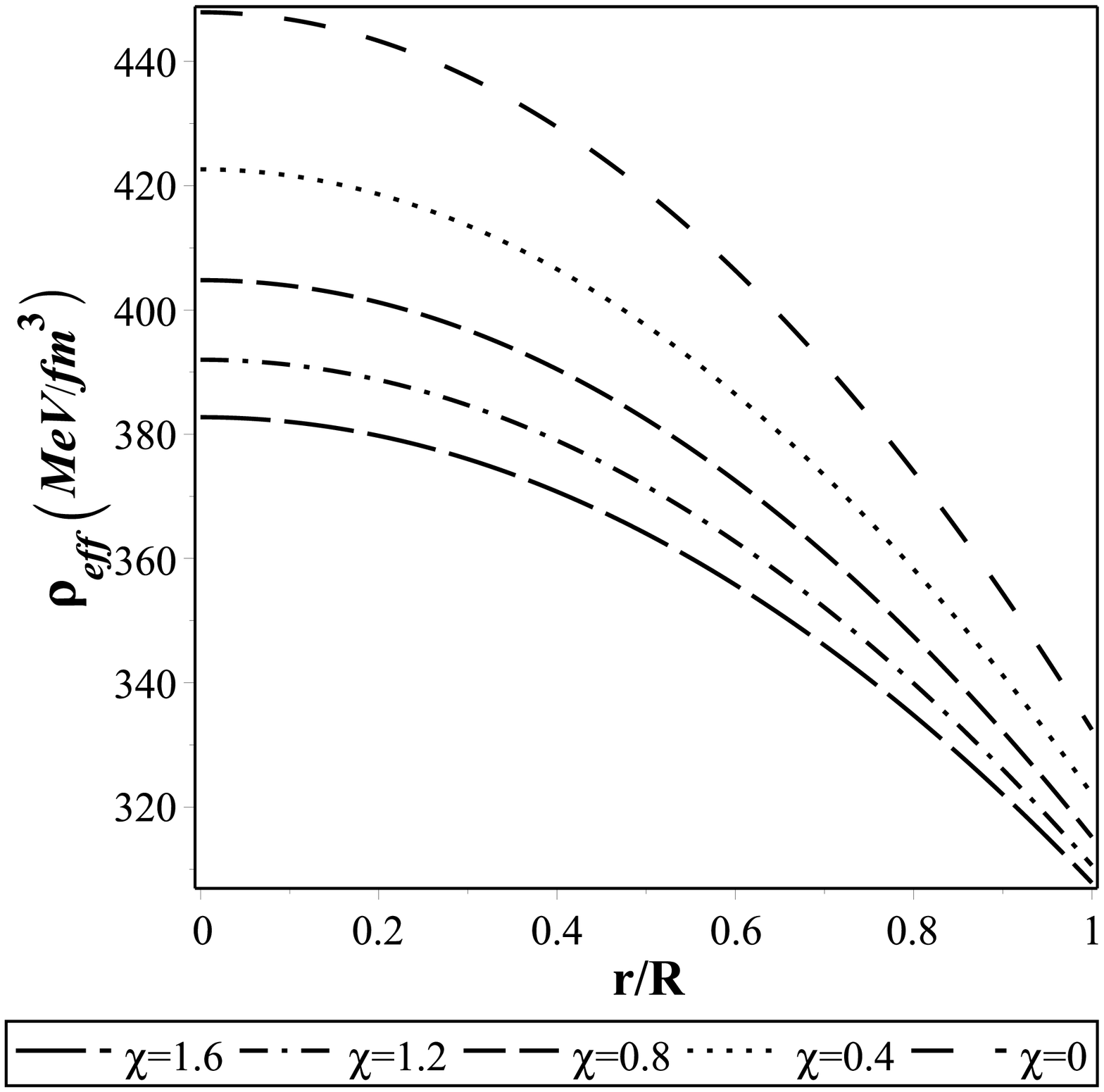}
\includegraphics[width=6cm]{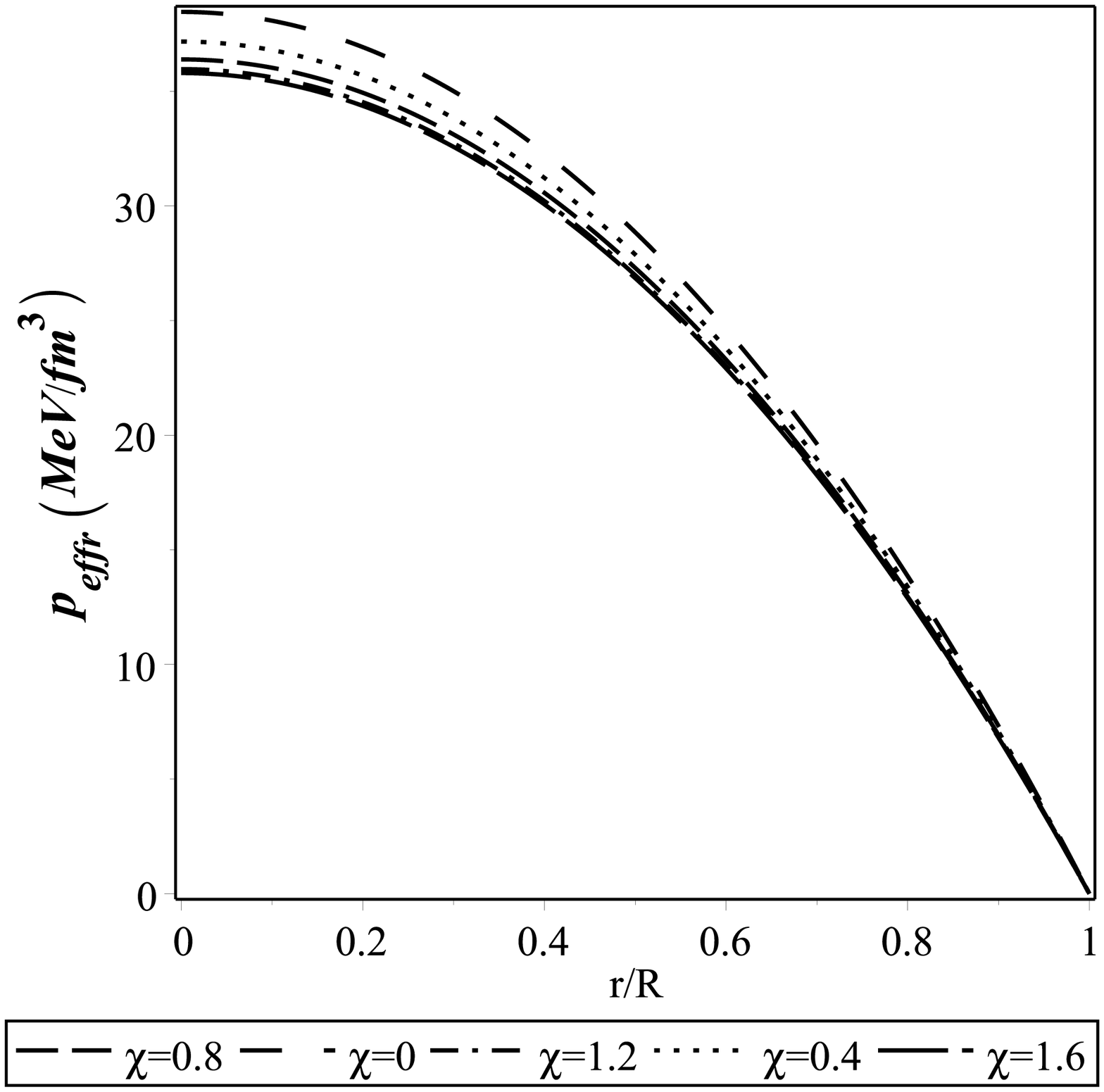}
\includegraphics[width=6cm]{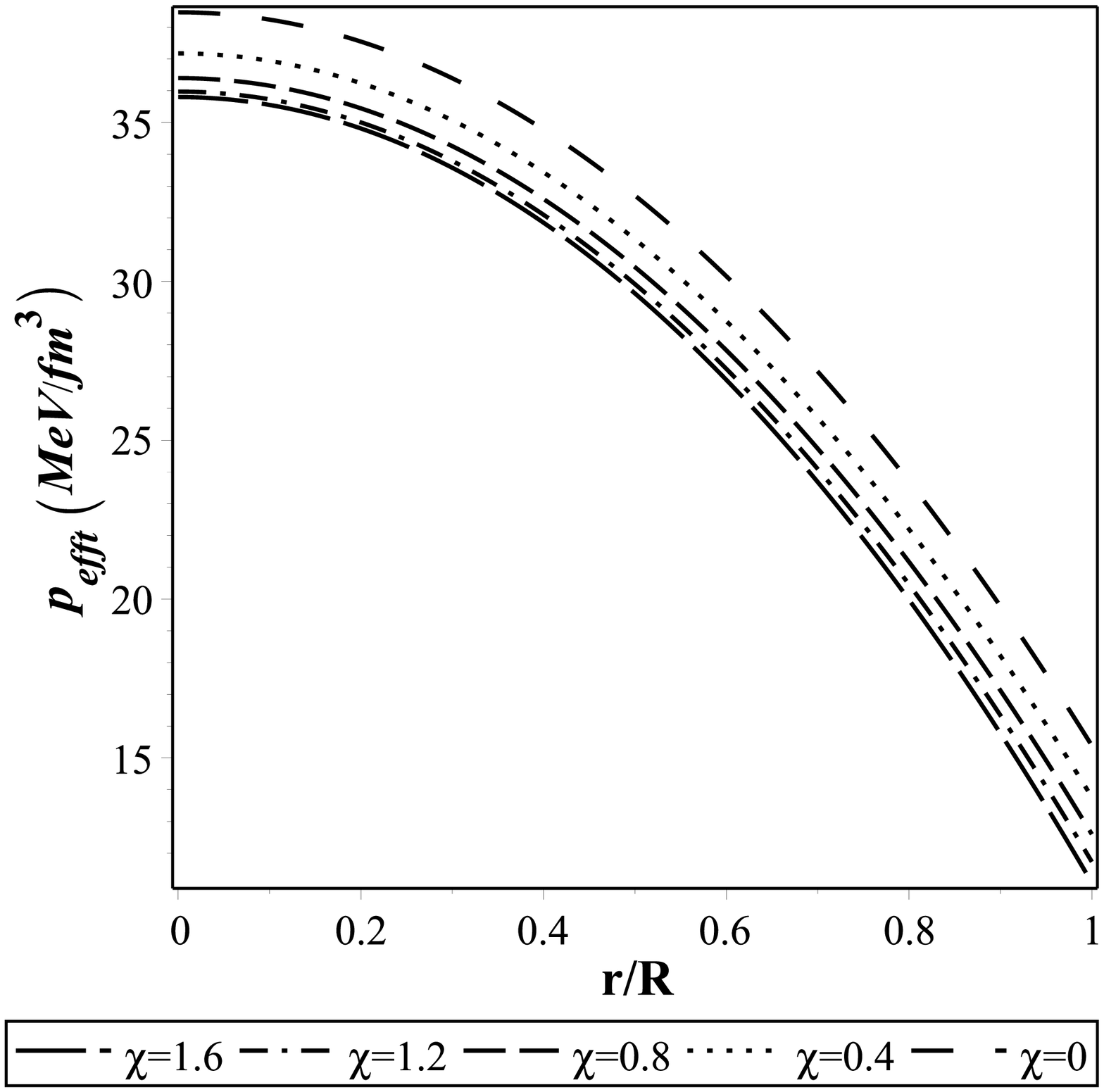}
\caption{Variation of i) ${\rho}_{\it eff}$ (upper panel), (ii) $p_{{{\it effr}}}$ (middle panel) and (iii) $p_{{{\it efft}}}$ (lower panel) as a function of the radial coordinate $r/R$ for the strange star $LMC~X-4$.} \label{Fig2}
\end{figure}


The variation of the physical parameters, viz., ${e}^{\lambda}$,~${e}^{\nu}$,~${\rho}_{{{\it eff}}}$,~${p}_{{{\it effr}}}$~and~${p}_{{{\it efft}}}$ with respect to the radial coordinate ($r/R$) in the framework of $f\left( R,\mathcal{T}\right)$ gravity theory are shown in Figs.~\ref{Fig1}~and~\ref{Fig2}.

The anisotropy~$\left(\Delta\right)$ for our system reads as
\begin{equation}\label{3.6}
\Delta=\frac { \left( c_{{1}}-\frac{1}{3}\right)  \left( \pi +\frac{\chi}{4} \right)  \left( \rho_{{0}}-\rho_{{c}} \right) {r}^{2}}{\pi \,{R}^{2}}.
\end{equation}


\begin{figure}[!htpb]\centering
	\includegraphics[width=6cm]{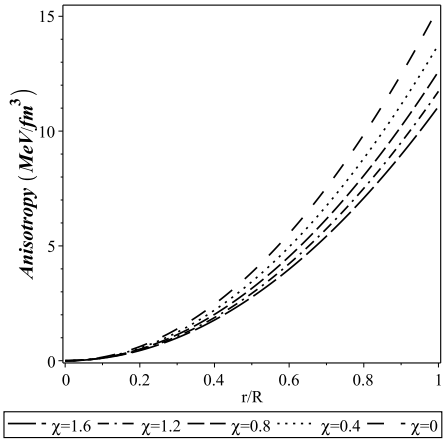}
			\caption{Variation of anisotropy as a function of the radial coordinate $r/R$ for the strange star $LMC~X-4$.} \label{Fig3}
\end{figure}


The variation of anisotropy with respect to radial coordinate $r/R$ is featured in Fig.~\ref{Fig3}. We find here in $f(R,\mathcal{T})$ gravity model that the anisotropy for the strange star is minimum at the center and maximum on the surface as prediction by Deb et al.~\cite{Deb2017} in the case of GR. 

The modified form of the energy conservation equation for the stress-energy tensor in the framework of $f\left(R,\mathcal{T}\right)$ can be written explicitly from Eq.~(\ref{1.6}) as
\begin{eqnarray}\label{3.7}
&\qquad\hspace{-4cm} -p_r^{{\prime}}-\frac{1}{2}\nu^{{\prime}} \left( \rho+p_r \right)+\frac{2}{r}\left({p_t}-{p_r}\right)\nonumber\\
&\qquad\hspace{1cm} -{\frac {\chi}{8\,\pi +2\,\chi}}\left(\rho^{{\prime}}+p_r^{{\prime}} +2p_t^{{\prime}} \right)=0.
\end{eqnarray}

Now using Eqs. (\ref{2.3}),~(\ref{2.11a}),~(\ref{2.14})~and~(\ref{3.7}) we find the hydrostatic equation for an anisotropic spherically symmetric compact stars in the framework of $f\left(R,\mathcal{T}\right)$ theory of gravity as follows
\begin{eqnarray}\label{3.8}
&\qquad p_r^{\prime}=-\Big[\big\lbrace 4 \pi {r}^{2}p_{{r}}+{\frac {m}{r}}+\frac{1}{2}\chi \left( \rho+3 p_{{r}}+2 p_{{t}} \right) {r}^{2} \big\rbrace \nonumber\\
&\qquad \left( \rho+p_{{r}} \right) -2 \left( p_{{t}}-p_{{r}} \right)  \left( 1-{\frac {2m}{r}} \right)\Big]\Big/\Big[ r \left( 1-{\frac {2m}{r}} \right)\nonumber\\
&\qquad  \big\lbrace 1+\frac {\chi}{8 \pi +2 \chi}\big[ 1+ {\frac {{\rm d}\rho}{{\rm d}p_{{r}}}}\left( 1+2 c_{{1}} \right)  \big] \big\rbrace\Big],
\end{eqnarray}
where we assume that SQM density parameter $\rho$ depends on it's radial pressure ${p}_{r}$ as ${\rho}={\rho} \left({{p}_{r}}\right)$. For ${\chi}=0$ Eq.~(\ref{3.8}) reduces to the standard form of the TOV equation as found in GR. Now using Eqs. (\ref{2.10}) and (\ref{2.11}) and also  considering Bag constant $B=83~MeV/{{fm}^{3}}$~\cite{Rahaman2014} with ${c_1}={0.2}$ we obtain exact solution of the Eq.~(\ref{3.8}). Here, using the observed values of the mass of different strange stars as presented in Table~\ref{Table 2} we can predict radii of the strange stars.

\section{Physical properties of the anisotropic stellar system in $f\left(R,\mathcal{T}\right)$ theory of gravity}\label{sec4}
In this section we shall test physical validity of the obtained solutions in the framework of $f \left(R,\mathcal{T}\right)$ theory of gravity. To this end, we study the energy conditions, Herrera cracking concept, adiabtic index, etc., in the following subsections.

\subsection{Energy conditions in $f\left(R,\mathcal{T}\right)$ gravity}\label{subsec4.1}
Our system will be consistent with the energy conditions, viz., the null energy condition (NEC), weak energy condition (WEC), strong energy condition (SEC) and dominant energy condition (DEC) only if it satisfy the following inequalities simultaneously, given as~\cite{SC2013}
\begin{eqnarray}\label{4.1.1}
&\qquad\hspace{-1cm} NEC:{\rho}_{eff}+p_{effr}\geq 0,~{{\rho}_{eff}}+{p_{efft}}\geq 0, \\ \label{4.1.2}
&\qquad\hspace{-1.2cm} WEC: {{\rho}_{eff}}+p_{effr}\geq 0,~{{\rho}_{eff}}\geq 0,~{{\rho}_{eff}}+{p_{efft}}\geq 0, \nonumber \\ \\ \label{4.1.3}
&\qquad\hspace{-1.4cm} SEC: {{\rho}_{eff}}+p_{effr}\geq 0,~{{\rho}_{eff}}+p_{efft}\geq 0,\nonumber \\
&\qquad\hspace{2cm} ~{{\rho}_{eff}}+{p_{effr}}+2 {p_{efft}}\geq 0, \\ \label{4.1.4}
&\qquad\hspace{-1.4cm} DEC: {{\rho}_{eff}}\geq 0,~{{{\rho}_{eff}}-{p_{effr}}}\geq 0,~{{{\rho}_{eff}}-{p_{efft}}}\geq 0. 
\end{eqnarray}

\begin{figure}[!htpb]
\centering
\includegraphics[width=4.5cm]{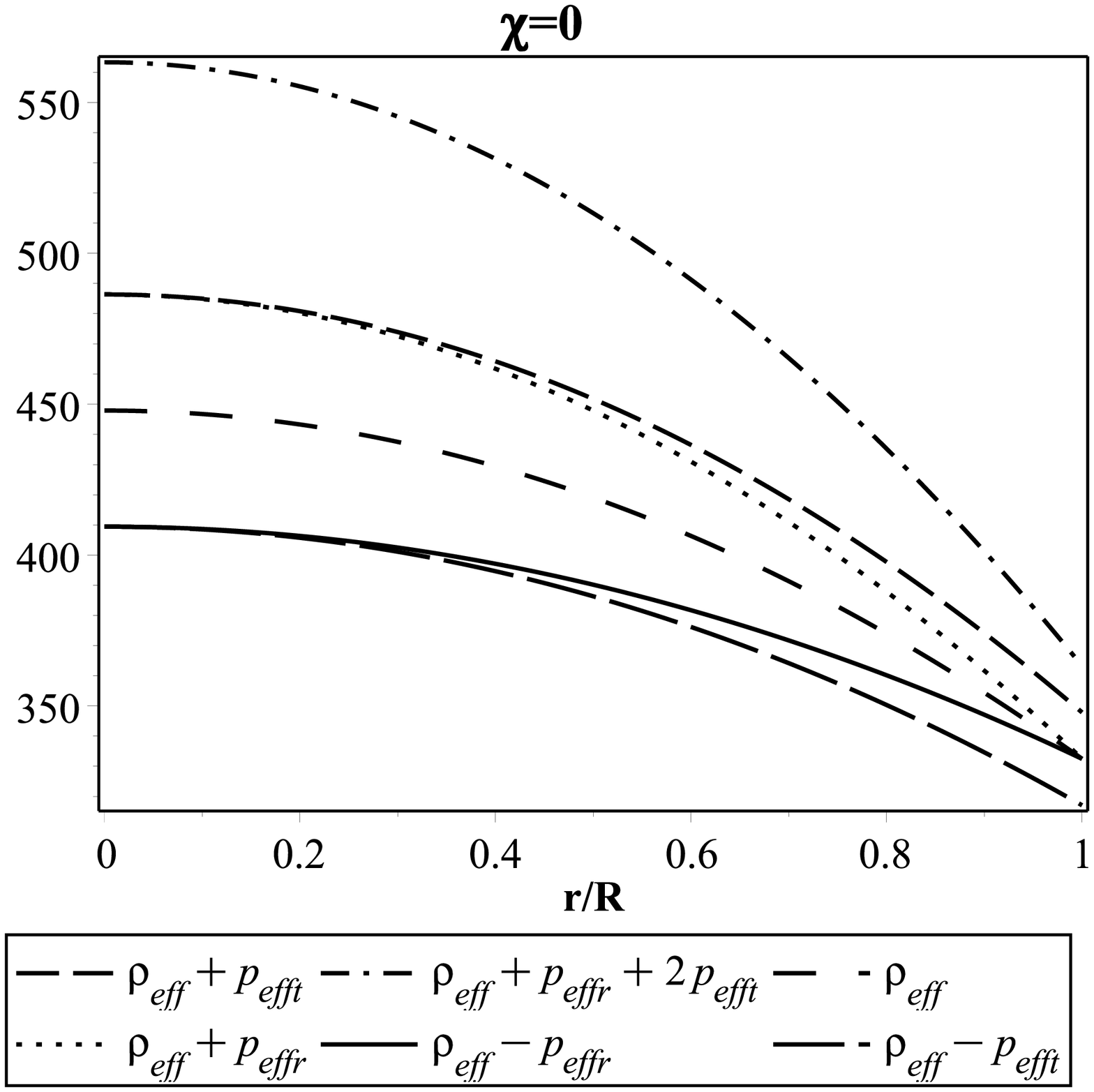}
\includegraphics[width=4.5cm]{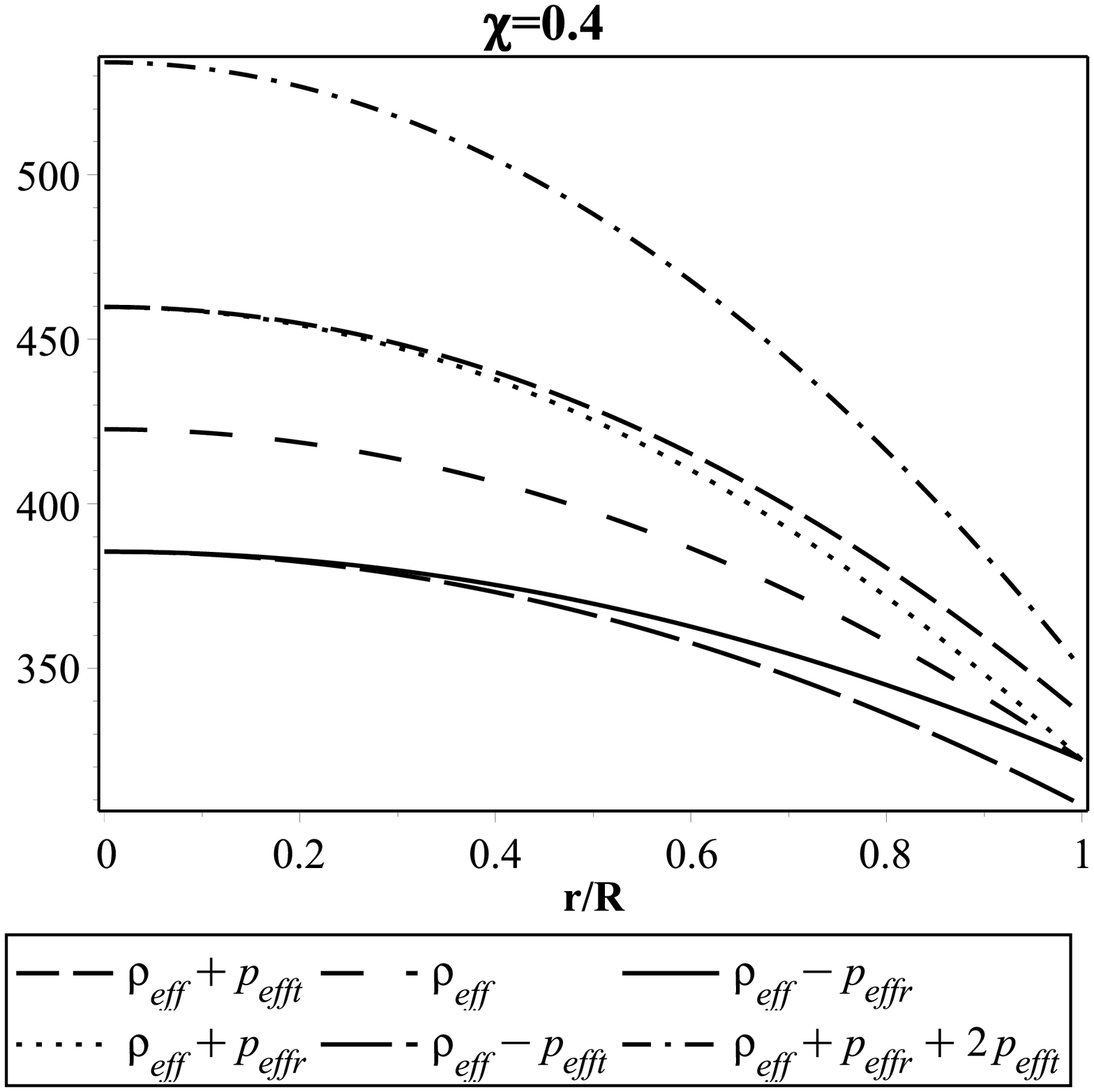}
\includegraphics[width=4.5cm]{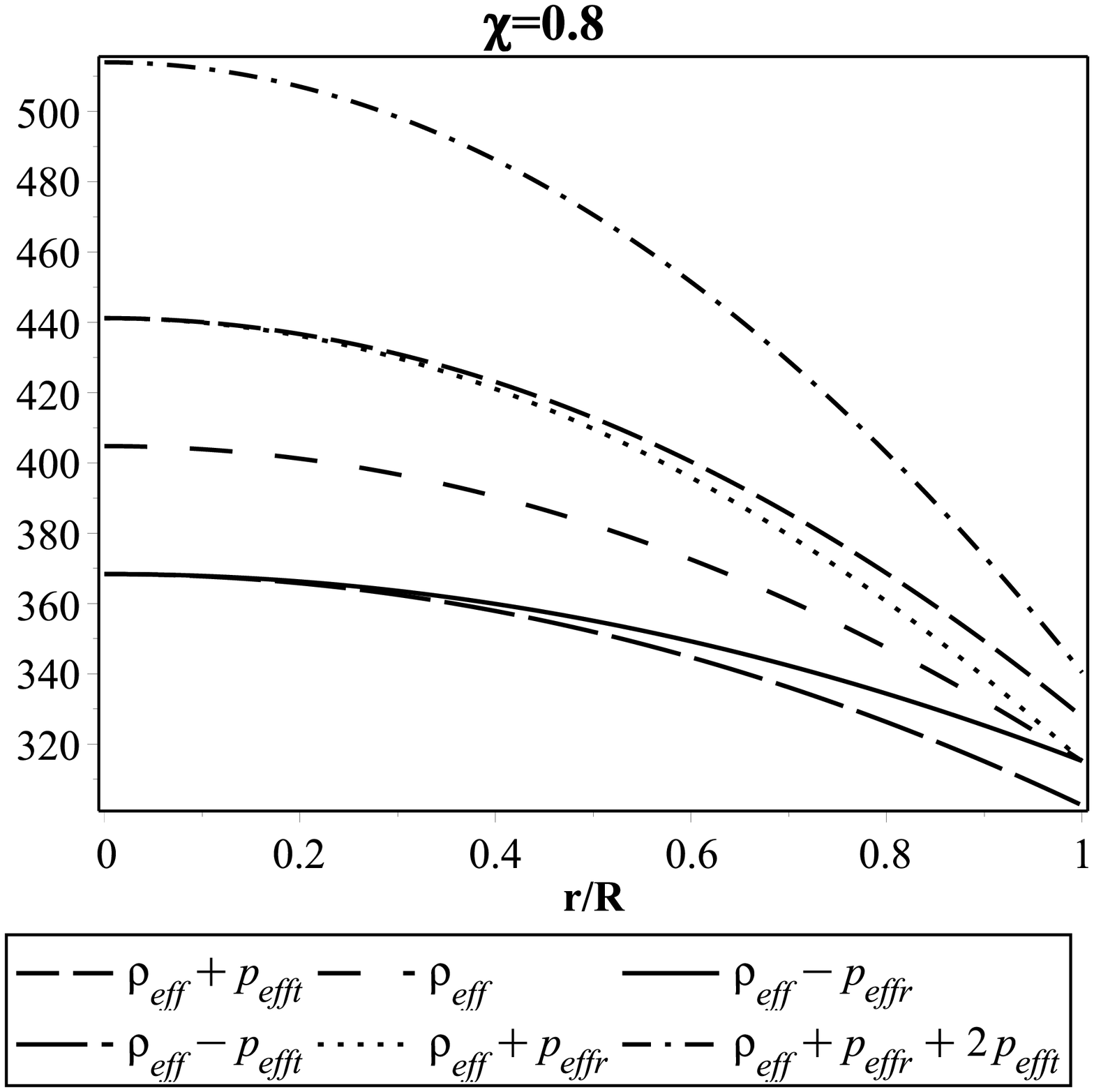}
\includegraphics[width=4.5cm]{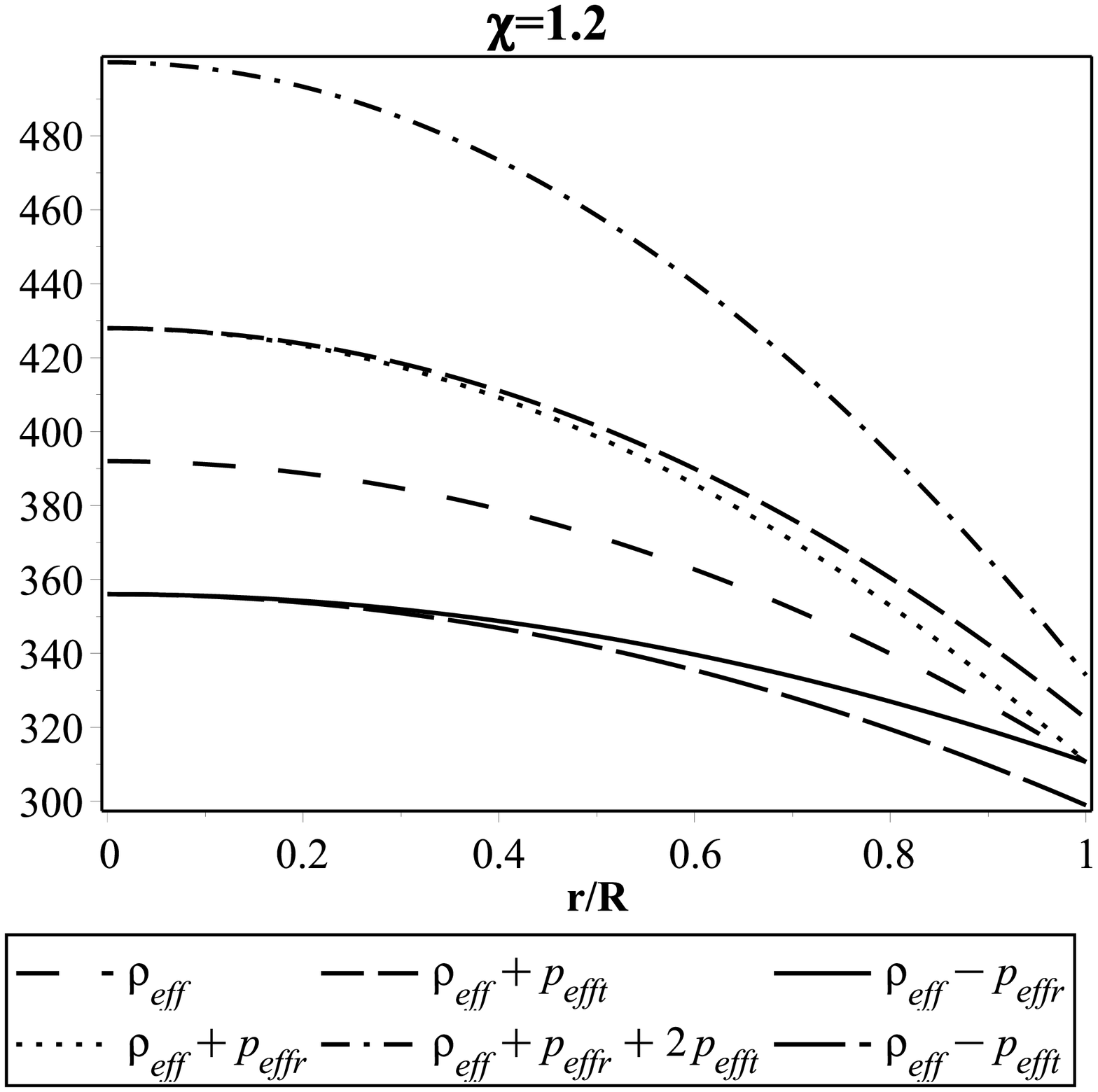}
\includegraphics[width=4.5cm]{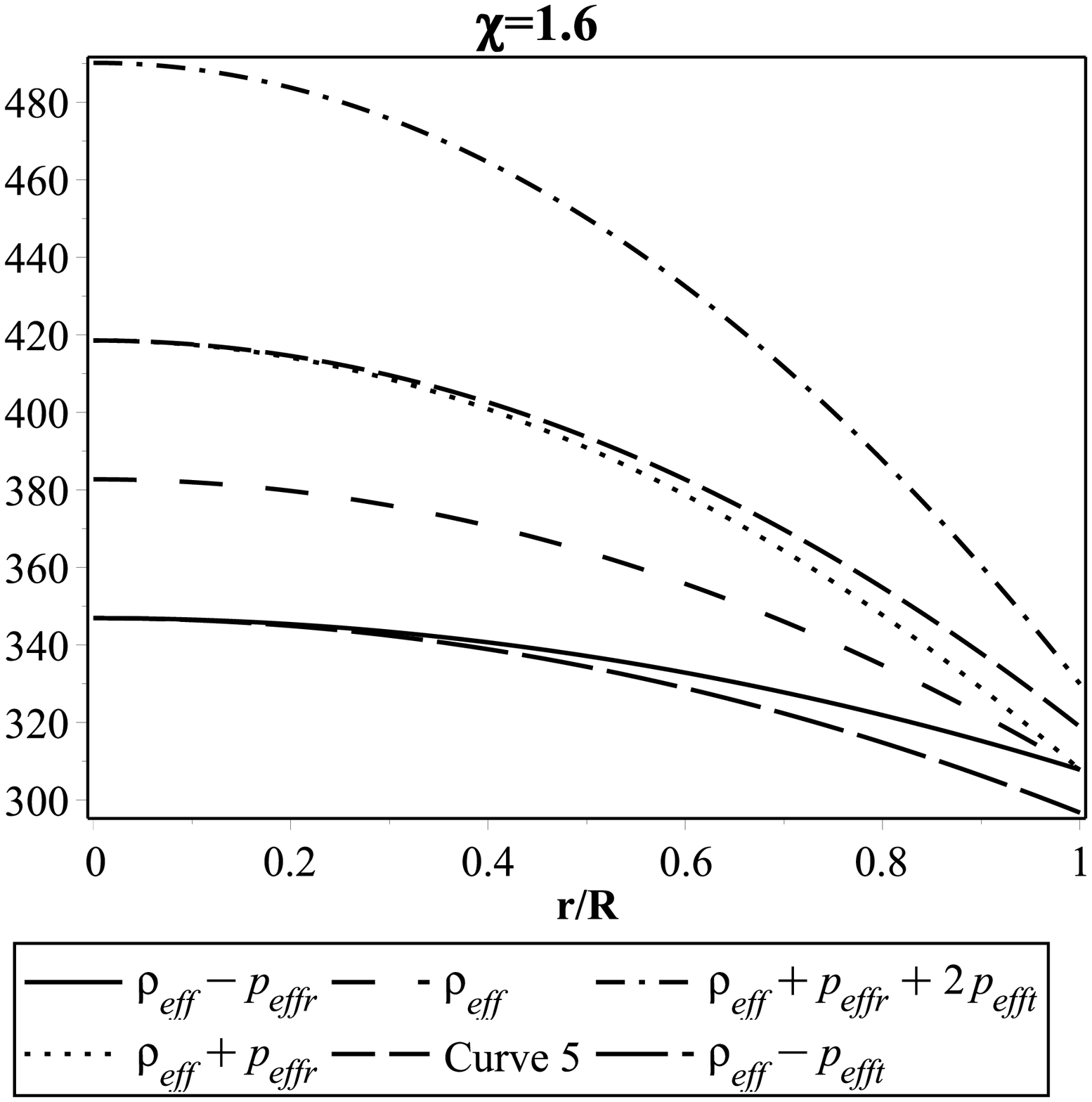}
\caption{Variation of energy conditions with the radial coordinate $r/R$ for $LMC\,X-4$ due to different chosen values of $\chi$.} \label{Fig4}
\end{figure}

The variation of the energy conditions with respect to the radial coordinate $r/R$ for the parametric values of $\chi$ are presented in Fig.~\ref{Fig4}, which clearly suggests that our system is consistent with all the energy conditions.

\subsection{Mass-radius relation}\label{subsec4.2}
\begin{figure}[!htpb]\centering
\includegraphics[width=8cm]{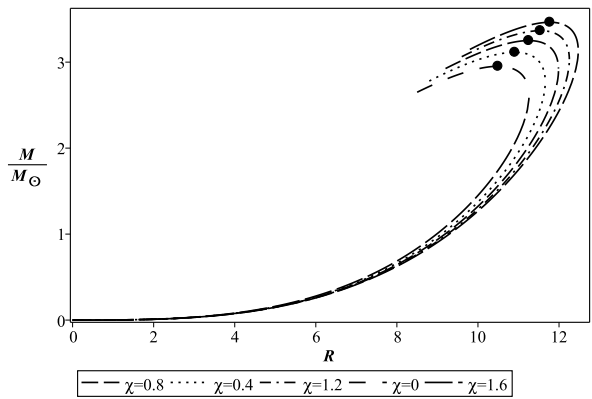}
	\caption{Mass~$(M/{M_{\odot}})$ vs Radius~($R$~in km) curve for the strange stars due to the different values of $\chi$. The solid circles are representing the maximum mass points for the strange stars.} \label{Fig5}
\end{figure}

Substituting Eqs.~(\ref{2.6}),~(\ref{2.10})~and~(\ref{2.11a}) into Eq.~(\ref{2.12}) the mass function for the present system is given as
\begin{eqnarray}\label{4.2.1}
\hspace{-0.5cm} m=\tilde{m} \Big\lbrace 1+{\frac {\chi}{12\pi }}\left( 1-3 c_{{1}} \right) \Big\rbrace +\frac{2}{9}\chi {r}^{3} \left( B-\frac{3}{2}c_{{2}} \right),
\end{eqnarray}
where $\tilde{m}=4 \pi \int_{0}^{r}\!{{\rho}} \left( r \right) {r}^{2}{dr}$ is the mass function of the SQM fluid distribution. For $\chi=0$ Eq.~(\ref{4.2.1}) reduce to the mass function $m(r)=\tilde{m}(r)$ as achieved in GR. Hence, clearly for $\chi \neq 0$ the coupling between matter and curvature terms produces a new kind of matter distribution having mass given as $m_{new}=\chi \lbrace {\frac {\tilde{m}}{12\pi }}\left( 1-3 c_{{1}} \right)  +\frac{2}{9}{r}^{3} \left( B-\frac{3}{2}c_{{2}} \right) \rbrace$. We have shown variation of the total mass~($M$), normalized in solar masses~($M_{\odot}$), with respect to the total radius ($R$) in Fig.~\ref{Fig5} for different values of $\chi$ and for a specific value of the bag constant as $B=83~ MeV/{{fm}^3}$~\cite{Rahaman2014}. Fig.~\ref{Fig5} shows that the mass-radius relation for the strange stars in $f\left(R,\mathcal{T}\right)$  gravity has achieved typical behaviour as in GR. Also, we find that for the chosen increasing values of $\chi$, i.e., $\chi=0$,~$0.4$,~$0.8$,~$1.2$~and~$1.6$ the values of the maximum masses are increasing gradually.


\begin{figure}[!htpb]
\centering
\includegraphics[width=6cm]{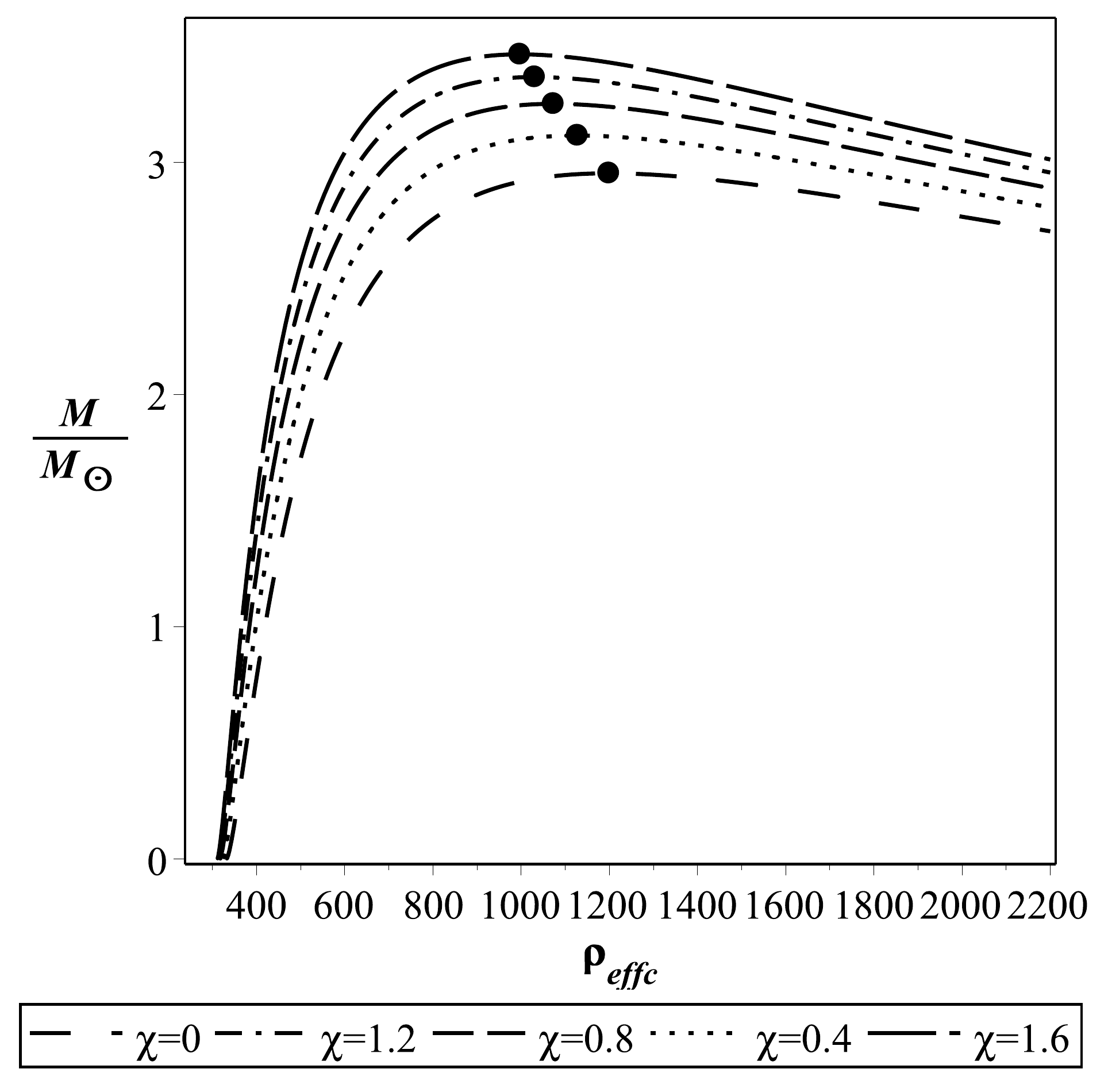}
\includegraphics[width=6cm]{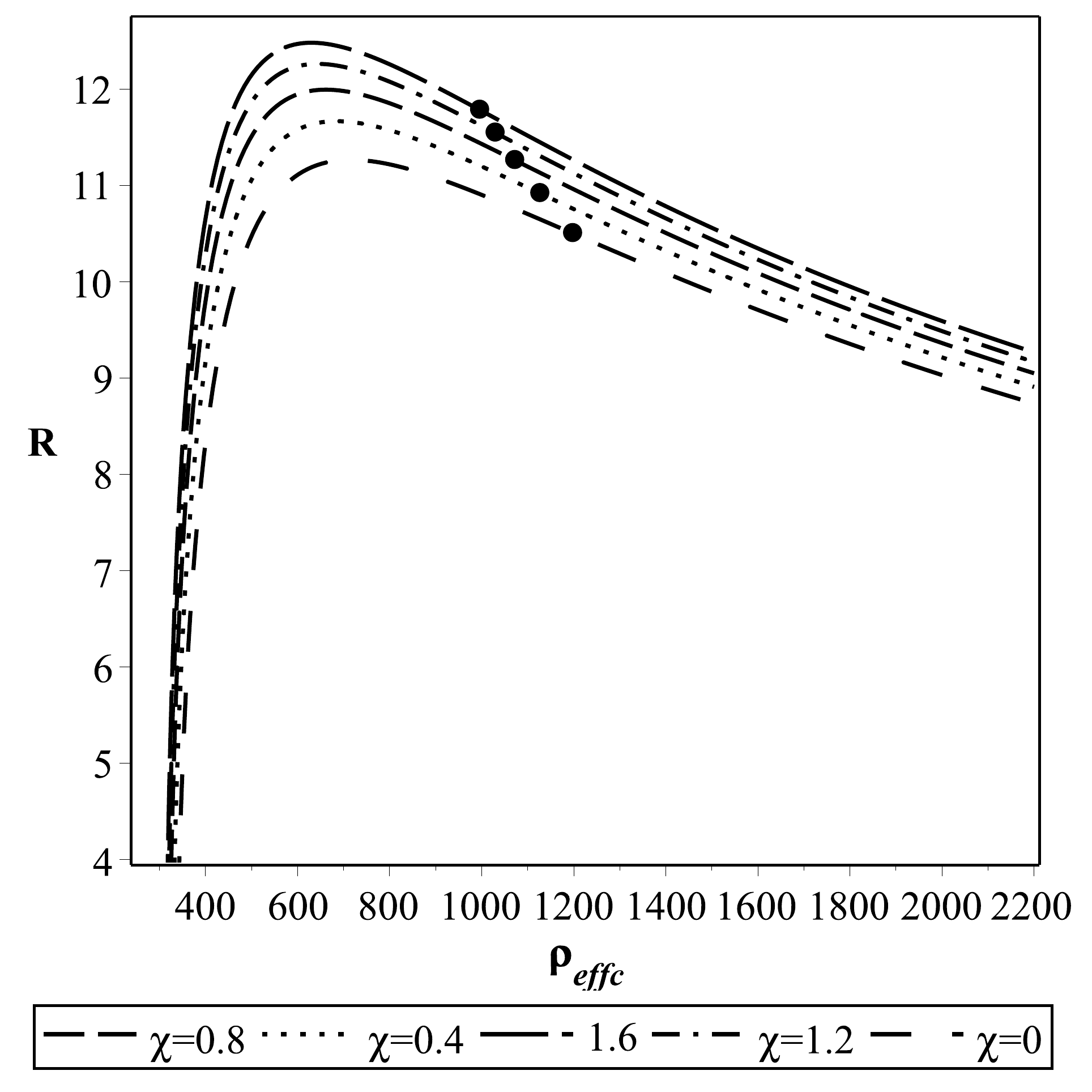}
\caption{Variation of (i) $M/{M_{\odot}}$ (left panel) and (ii) $R$~in $km$ (right panel) as a function of the central density $({{\rho}_{effc}}~in~MeV/{fm}^3)$ due to different values of $\chi$. The solid circles are representing the maximum mass points for the strange stars.} \label{Fig6}
\end{figure}

In Fig.~\ref{Fig6} the variation of $M$, normalized in $M_{\odot}$ and the variation of $R$ with respect to the central density, ${\rho}_{effc}$ are shown in the upper and lower panel, respectively. The upper panel of Fig.~\ref{Fig6} features that as the values of $\chi$ increases the maximum mass points are achieved for the lower values of $({\rho}_{effc})$. We find, for $\chi=0$ the maximum mass $M_{max}=2.951~{{M}_{\odot}}$ is obtained for ${\rho}_{effc}=2.14\times {{10}^{15}}~gm/{{cm}^3}$. On the other hand, for $\chi=1.6$ the maximum mass increases to the value $M_{max}=~3.464 {{M}_{\odot}}$ and the corresponding value of the central density decreases to ${\rho}_{effc}=1.78\times {{10}^{15}}~gm/{{cm}^3}$. The lower panel of Fig.~\ref{Fig6} presents that as the value of $\chi$ increases the value of the radius increases gradually. We find, for $\chi=0$ the radius corresponding to the maximum mass point is $R_{Mmax}=10.498~km$ and as $\chi$ increases to the value $\chi=1.6$ the radius corresponding to the maximum mass point also increases to the value $R_{Mmax}=11.779~km$. Hence, as the value of $\chi$ increases both the mass and the radius of the strange stars increase and the stars become less compact.

\subsection{Stability of the stellar model}\label{subsec4.3}
To discuss stability of the stellar model we shall study (i) Modified form of the TOV equation in $f\left(R,\mathcal{T}\right)$ gravity, (ii) Herrera cracking concept and (iii) Adiabatic index in the following sub-subsections.

\subsubsection{Modified form of the TOV equation in $f\left(R,\mathcal{T}\right)$ gravity}\label{subsubsec4.3.1}
We have presented the modified form of the energy conservation equation for the stress-energy tensor in the framework of $f\left(R,\mathcal{T}\right)$ theory of gravity in Eq.~(\ref{1.6}) and later we have shown it in a more concise form in Eq.~(\ref{1.7}). Hence the modified form of the Tolman-Oppenheimer-Volkoff (TOV) equation as already presented in Eq.~(\ref{3.7}) is given as
\begin{eqnarray*}
&\qquad\hspace{-4cm} -p_r^{{\prime}}-\frac{1}{2}\nu^{{\prime}} \left( \rho+p_r \right)+\frac{2}{r}\left({p_t}-{p_r}\right)\nonumber\\
&\qquad\hspace{1cm} -{\frac {\chi}{8\,\pi +2\,\chi}}\left(\rho^{{\prime}}+p_r^{{\prime}} +2p_t^{{\prime}} \right)=0,
\end{eqnarray*}
where the first term represents the hydrodynamic force~$(F_h)$, the second term denotes gravitational force~$F_g$ and the third term indicates anisotropic force~$(F_a)$. Here, the last term is the resultant of the coupling between the matter and the geometry and we are introducing it as the 'modified force'~$F_m$. Hence, the modified TOV equation predicts that in $f\left(R,\mathcal{T}\right)$ gravity also, sum of all the forces are zero, i.e., ${F_h}+{F_g}+{F_a}+{F_m}=0$. So, in terms of equilibrium of the forces our system is completely stable. Clearly, for $\chi=0$ the extra force term~$F_m$ will be zero and the usual form of the TOV equation as in GR will be retrieved. 

\begin{figure}[!htpb]
\centering
    \includegraphics[width=4.5cm]{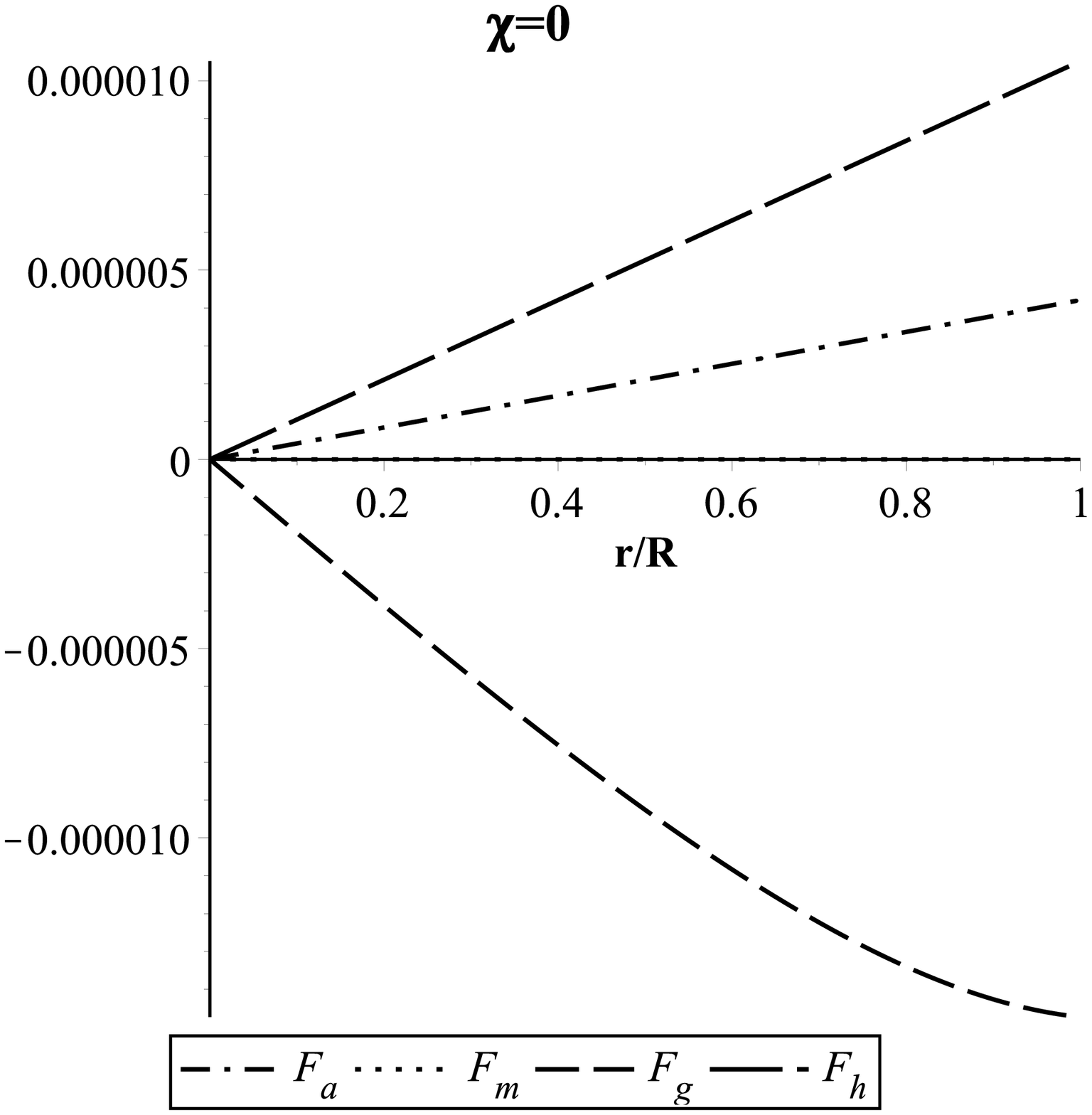}
   \includegraphics[width=4.5cm]{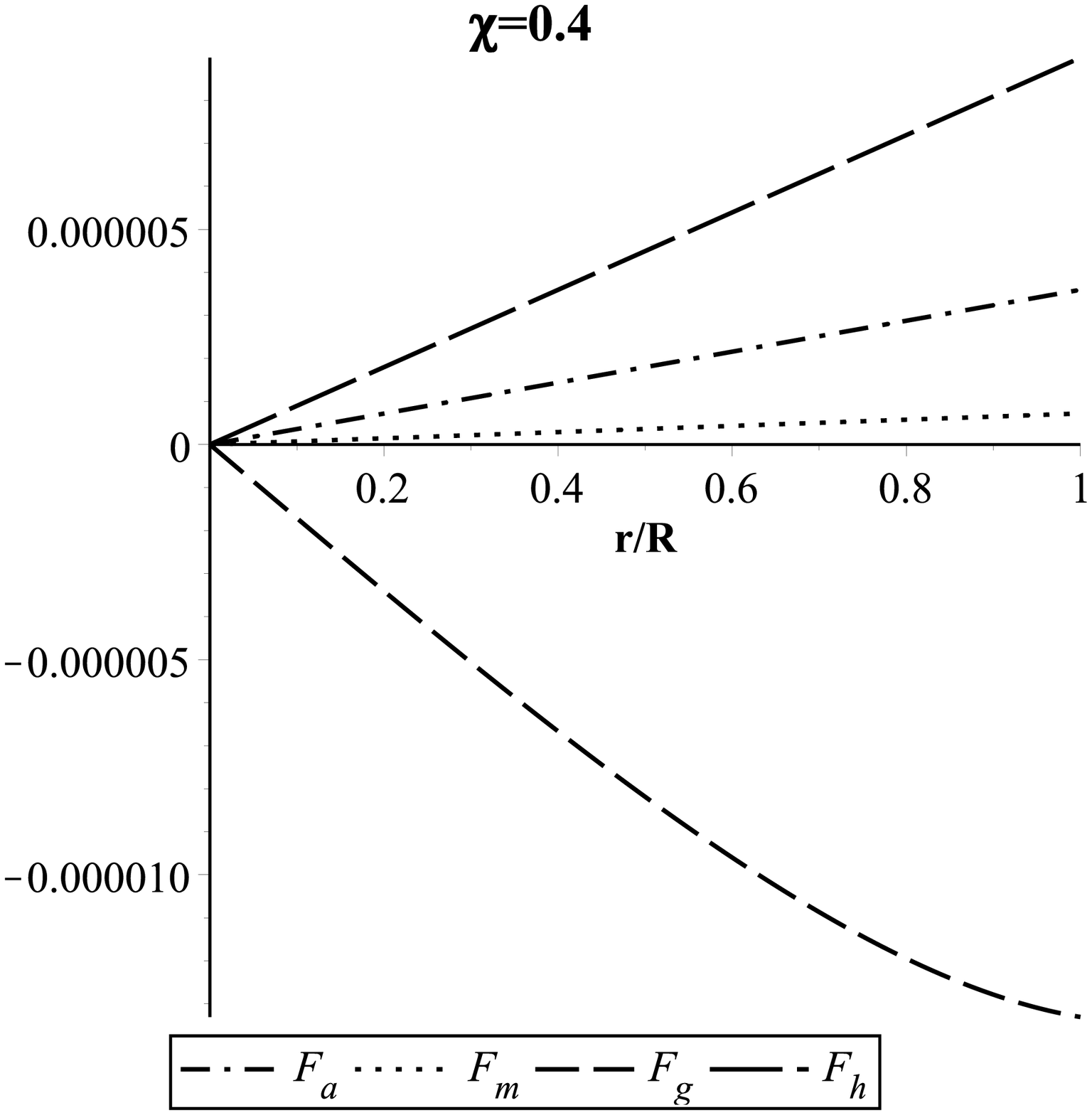}
    \includegraphics[width=4.5cm]{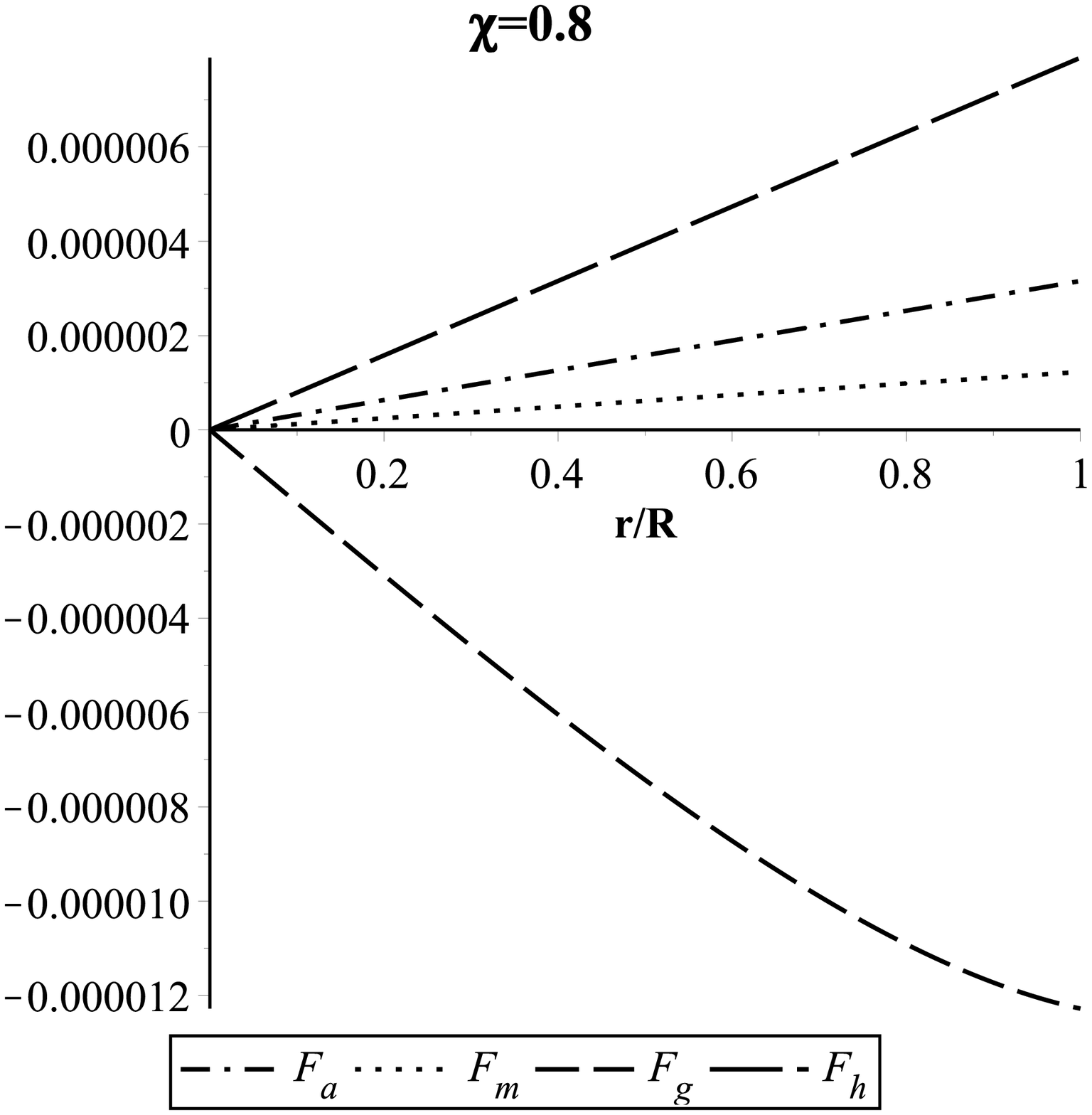}
   \includegraphics[width=4.5cm]{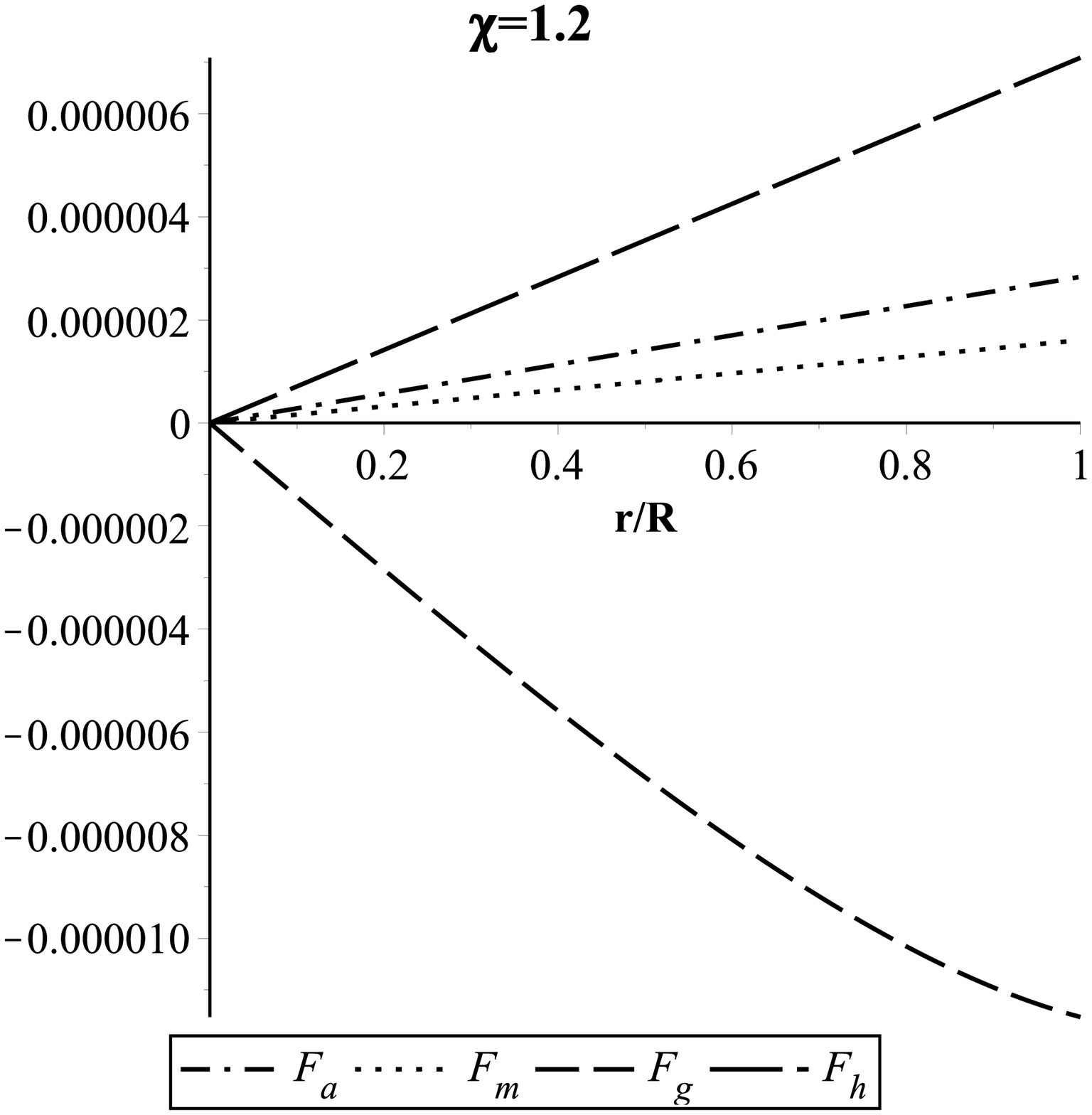}
   \includegraphics[width=4.5cm]{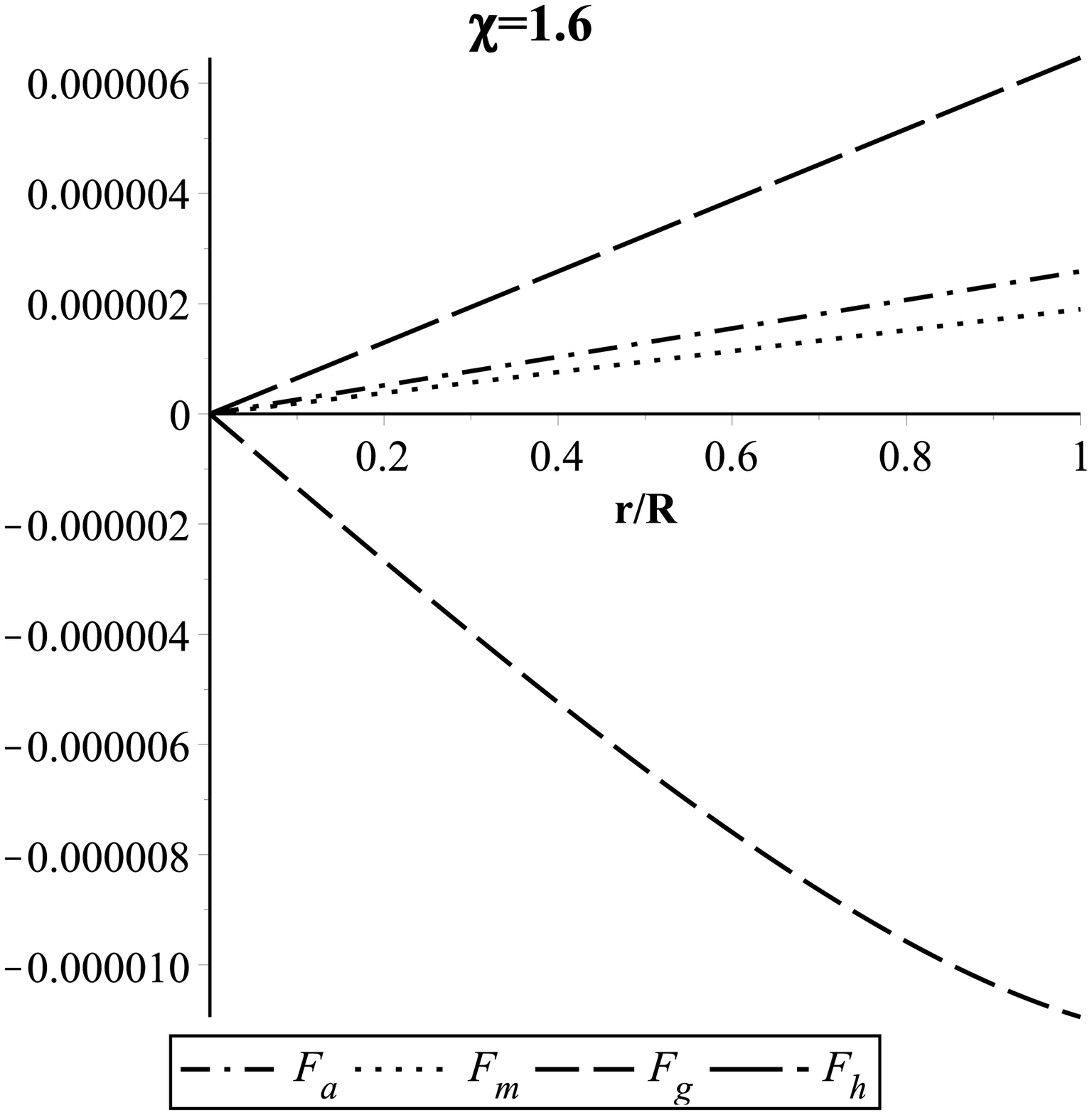}
   \caption{Variation of the different forces with respect to the radial coordinate $r/R$ for $LMC\,X-4$ due to different chosen values of $\chi$.} \label{Fig7}
\end{figure}

In Fig.~\ref{Fig7} we have shown variation of the different forces against the radial coordinate $r/R$ due to different chosen values of $\chi$. We find that the equilibrium of the forces is achieved due to all the values of $\chi$ and confirms stability of the system. Fig.~\ref{Fig7} features that the inward pull of $F_g$ is counter balanced by the combined effect of $F_h$,~$F_a$~and~$F_m$ which acts along the outward direction. Hence, we find that the nature of the modified force, $F_m$ is repulsive and acts along the outward directions.

\subsubsection{Herrera cracking concept}\label{subsubsec4.3.2}
To establish stability of the stellar system now we shall study the concept of Herrera's cracking. For a physically acceptable stellar system the causality condition must be satisfied, which demands that square of the radial $(v^2_{sr})$ and tangential $(v^2_{st})$ sound speeds should lie within the limit $\left[0,1\right]$, i.e., explicitly $0 \leq v^2_{sr} \leq 1$ and $0 \leq v^2_{st} \leq 1$. According to Herrera~\cite{Herrera1992} and Abreu~\cite{Abreu2007} for a physically stable stellar system made of anisotropic fluid distribution the difference of  square of the sound speeds should maintain it's sign inside the stellar system and specially for a potentially stable region square of the radial sound speed should be greater than the square of the tangential sound speeds. Hence, according to Herrera's cracking concept the required condition is  $|{v^2_{st}}- {v^2_{sr}}|\leq 1$. For our system $(v^2_{sr})$ and $(v^2_{st})$ are given as
\begin{eqnarray}\label{4.3.2.1}
&\qquad {v^2_{{{\it sr}}}}={\frac {3 \chi c_{{1}}+4 \pi +3 \chi}{-3 \chi c_{{1}}+12 \pi +\chi}}, \\ \label{4.3.2.2}
&\qquad {v^2_{{{\it st}}}}={\frac {2 \left(6 \pi  c_{{1}}+3 \chi c_{{1}}+\chi\right)}{-3 \chi c_{{1}}+12 \pi +\chi}}.
\end{eqnarray}

\begin{figure}[!htpb]
\centering
\includegraphics[width=6cm]{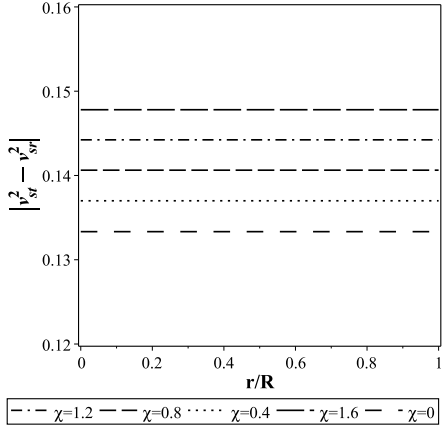}
\caption{Variation of $|{v^2_{st}}- {v^2_{sr}}|$ with the radial coordinate $r/R$ for $LMC\,X-4$.} \label{Fig8}
\end{figure}

We have featured variation of $|{v^2_{st}}- {v^2_{sr}}|$ with respect to the radial coordinate in Fig.~\ref{Fig8} and as $|{v^2_{st}}- {v^2_{sr}}|\leq 1$, so our system is consistent with the concept of Herrera's cracking, which again confirms the stability of our stellar system.

\subsubsection{Adiabatic Index}\label{subsubsec4.3.3}
The stability of both the relativistic and non-relativistic stars can be examined by studying adiabatic index $(\Gamma)$ of the system. For a given density it also can characterize the stiffness of the EOS. Following the pioneering work by Chandrasekhar~\cite{Chandrasekhar1964} several authors~\cite{Hillebrandt1976,Chan1994,Herrera1997,Horvat2010,Doneva2012,Silva2015} studied the dynamical stability of the stellar system against an infinitesimal radial perturbation. For a dynamically stable stellar system Heintzmann and Hillebrandt~\cite{Heintzmann1975} have shown that adiabatic indices should exceed $4/3$ inside the stellar system. Now the radial $({\Gamma}_r)$ and tangential $({\Gamma}_t)$ adiabatic indices can be defined as
\begin{eqnarray}\label{4.3.3.1}
& \qquad {\Gamma}_r=\frac{{p_{effr}}+{{\rho}_{eff}}}{{p_{effr}}}\,\frac{d{p_{effr}}}{d{{\rho}_{eff}}}=\frac{{p_{effr}}+{{\rho}_{eff}}}{{p_{effr}}}\,{v^2_{sr}},\\\label{4.3.3.2}
& \qquad {\Gamma}_t=\frac{{p_{efft}}+{{\rho}_{eff}}}{{p_{efft}}}\,\frac{d{p_{efft}}}{d{{\rho}_{eff}}}=\frac{{p_{efft}}+{{\rho}_{eff}}}{{p_{efft}}}\,{v^2_{st}}. 
\end{eqnarray}

\begin{figure}[!htp]
\centering
    \includegraphics[width=6cm]{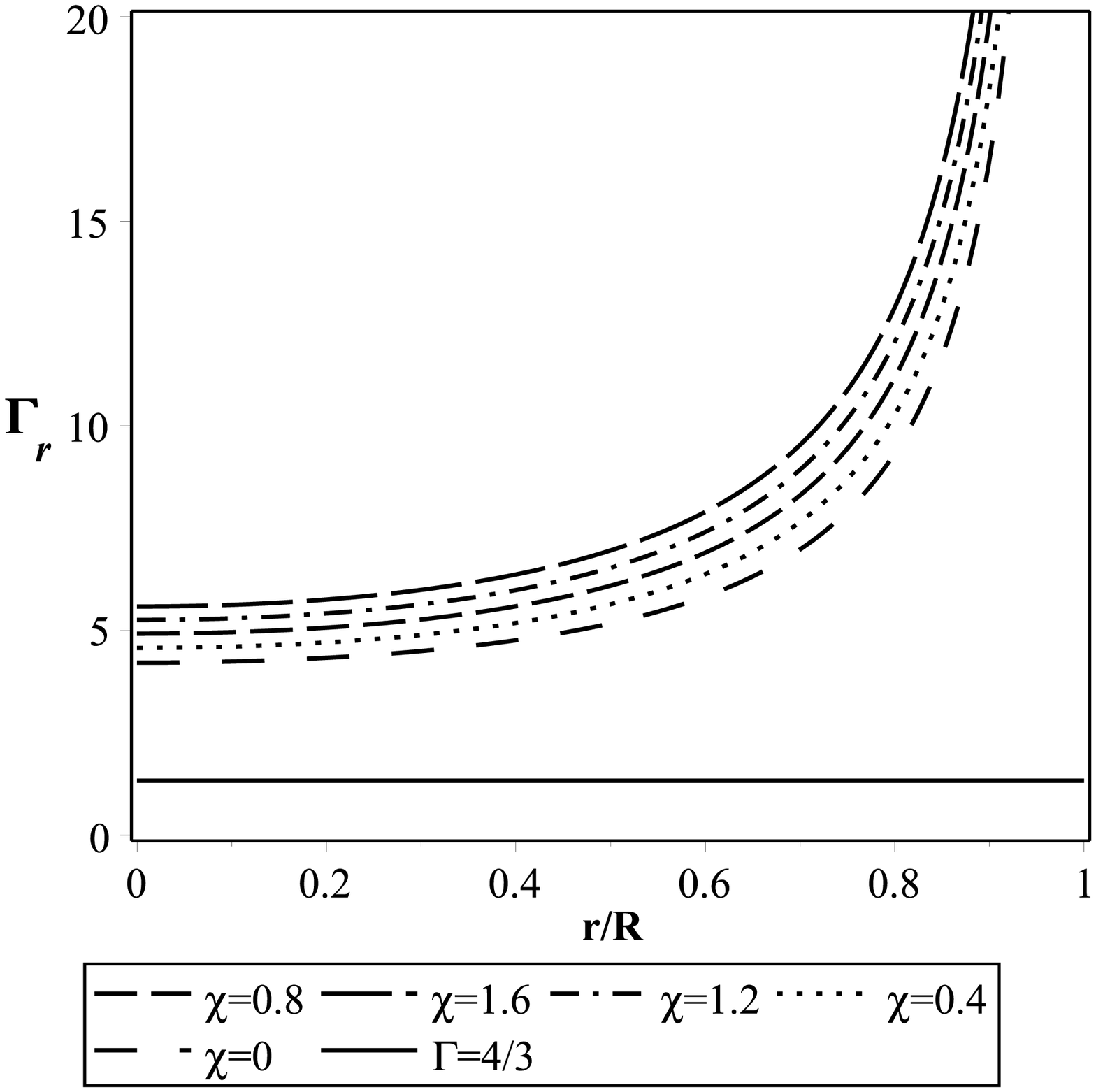}
    \includegraphics[width=6cm]{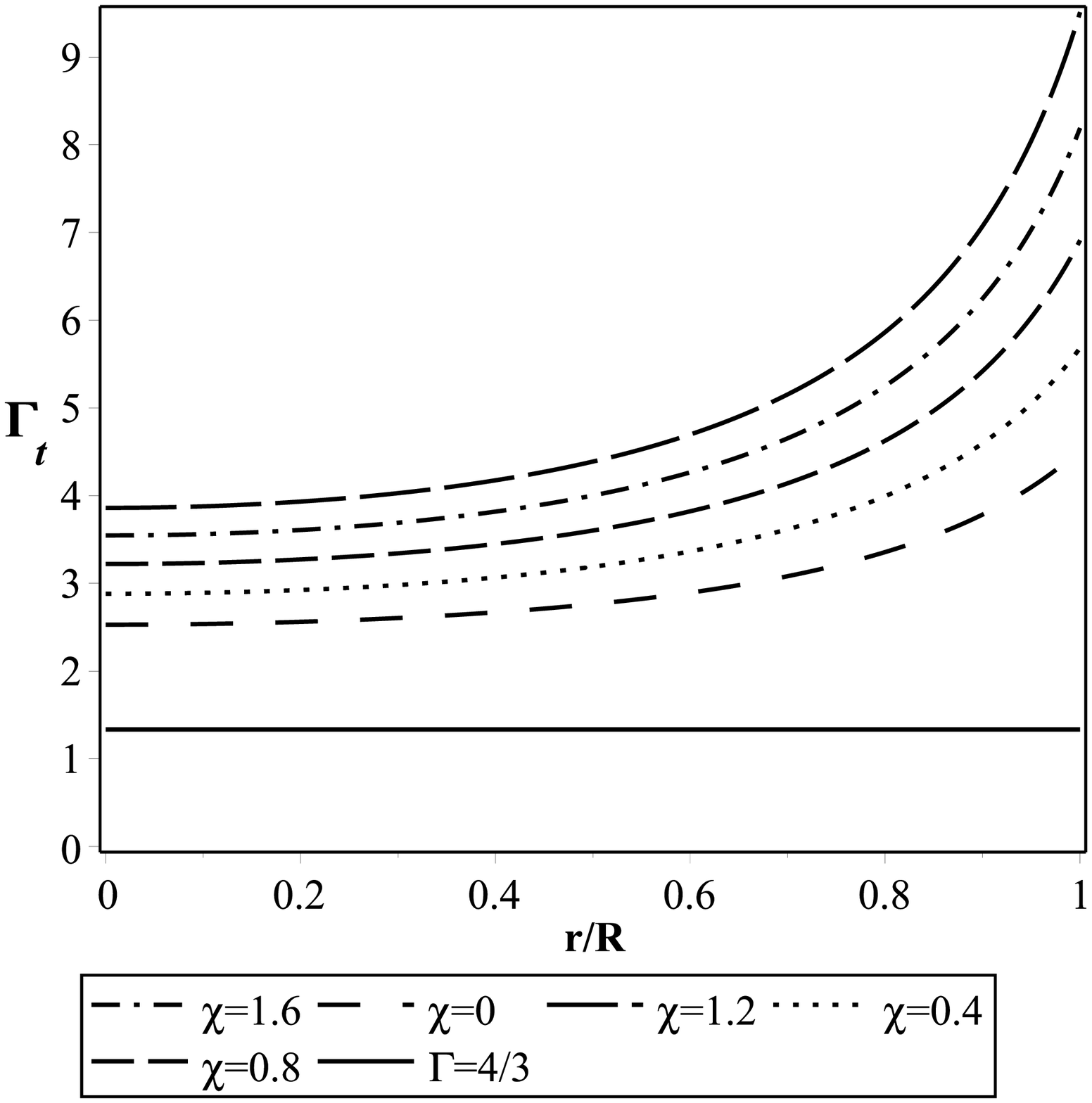}
    \caption{Variation of i) ${\Gamma}_r$ (upper panel) and ii) ${\Gamma}_t$ (lower panel) with the radial coordinate $r/R$ for $LMC\,X-4$.} \label{Fig9}
\end{figure}

In Fig.~\ref{Fig8} we have shown the variation of ${\Gamma}_r$ (upper panel) and ${\Gamma}_t$ (lower panel) against the radial coordinate $r/R$ which demonstrate that in both the cases the values of the adiabatic indices are greater than $4/3$ through out the system. Hence, our system is completely stable against the radial pulsations.

\subsection{Compactification factor and redshift}\label{subsec4.4}
The compactification factor~$(u)$ for our system is expressed by
\begin{eqnarray}\label{4.4.1}
&\qquad\hspace{-2cm} u(r)=\frac{m(r)}{r}=-\frac {1}{ \lbrace  \left( \frac{3}{8}c_{{1}}-\frac{1}{8}\right) {\chi}^{2}+\frac{9}{4}\pi\left( c_{{1}}+\frac{1}{9}\right) \chi+{\pi }^{2} \rbrace {R}^{5}}\nonumber \\
&\qquad\hspace{-2cm}\Big[ 8r^2  \Big\lbrace -\frac{1}{32}B \left( R^2-r^2 \right)  \left( c_{{1}}-\frac{1}{3}\right){R}^{3}{\chi}^{3}\nonumber \\
&\qquad\hspace{-2cm}+ \big[ -\frac{3}{16}B{R}^{3}  \left( R^2-r^2 \right)\left( c_{{1}}-1 \right) \pi -{\frac { \left( 3
c_{{1}}-1 \right) M{r}^{2}}{64}} \big] {\chi}^{2}\nonumber \\
&\qquad\hspace{-2cm}-\frac{1}{4}\pi \big\lbrace B \left( R^2-r^2 \right)  \left( c_{{1}}-\frac{10}{3}
 \right) {R}^{3}\pi +\frac{1}{16}\big[15 M \big\lbrace  \left( c_{{1}}+1 \right) {R}^{2}\nonumber \\
&\qquad\hspace{-2cm}+\frac{1}{5} \left( c_{{1}}-\frac{13}{3} \right) {r}^{2} \big\rbrace \big] \big\rbrace \chi+ \big[ \left( B{R}^{5}-B{R}^{3}{r}^{2} \right) \pi\nonumber \\
&\qquad\hspace{-2cm} -\frac {1}{16}\left( 5{R}^{2}-3{r}^{2} \right) M \big] {\pi }^{2}\Big\rbrace \Big].
\end{eqnarray}

Again, expression for the redshift function in the present model is given as
\begin{eqnarray}
&\qquad\hspace{-5cm} Z={e^{-{\nu(r)}/2}}-1\nonumber \\
&\qquad\hspace{-1cm} =exp\Bigg\lbrace -\frac {1}{73728\,\nu_{{4}} \left( -3\,\chi\,c_{{1}}+12\,\pi +\chi \right) } \Big[ \nu_{{3}}{\rm arctanh} \Big\lbrace\Big(\big[ 16\,\nu_{{2}}B{R}^{5}\nonumber\\
&\qquad\hspace{-1cm}-32\,B{R}^{3}{r}^{2}\nu_{{2}}+6\,M{r}^{2} \left(\pi +\chi \right)  \big] \lambda_{{2}}-5\,\lambda_{{1}}M\pi \,{R}^{2}\Big)/{16\nu_{{4}}}\Big\rbrace\nonumber\\
&\qquad\hspace{-1cm} -294912\big\lbrace\left( \frac{3}{8}c_{{1}}+\frac{1}{2} \right) \chi+\pi  \big\rbrace \nu_{{4}}\ln\Big[ 384\,{R}^{5}{r}^{2}\lambda_{{2}}\nu_{{2}}B\nonumber\\
 &\qquad\hspace{-1cm}-384\,\lambda_{{2}}{r}^{4}\nu_{{2}}B{R}^{3}-120{R}^{2}{r}^{2}\lambda_{{1}}M\pi +72\,M{r}^{4} \left( \pi +\chi \right) \lambda_{{2}}\nonumber\\
&\qquad\hspace{-1cm}+\nu_{{5}}\Big] -\nu_{{3}}{\rm arctanh}\Big[\big\lbrace -\lambda_{{2}}\nu_{{2}}B{R}^{5}-{\frac {5\,\lambda_{{1}}M\pi \,{R}^{2}}{16}}\nonumber\\
&\qquad\hspace{-1cm}+\frac{3}{8}\lambda_{{2}}{R}^{2}M \left( \pi +\chi \right)\big\rbrace/{\nu_{{4}}}\Big] +442368\,\nu_{{4}} \Big\lbrace\Big[ \frac{2}{3}\pi + \left( \frac{1}{4}c_{{1}}+\frac{1}{3}\right) \chi\Big]\nonumber\\
 &\qquad\hspace{-1cm} \ln\Big\lbrace 24\,{R}^{4}\nu_{{1}} \left(R-2M \right)  \Big\rbrace +\ln  \left( 1-{\frac {2M}{R}} \right) \lambda_{{2}} \Big\rbrace\Big] \Bigg\rbrace-1.\nonumber\\
\end{eqnarray}


\begin{figure}[!htpb]\centering
\includegraphics[scale=.3]{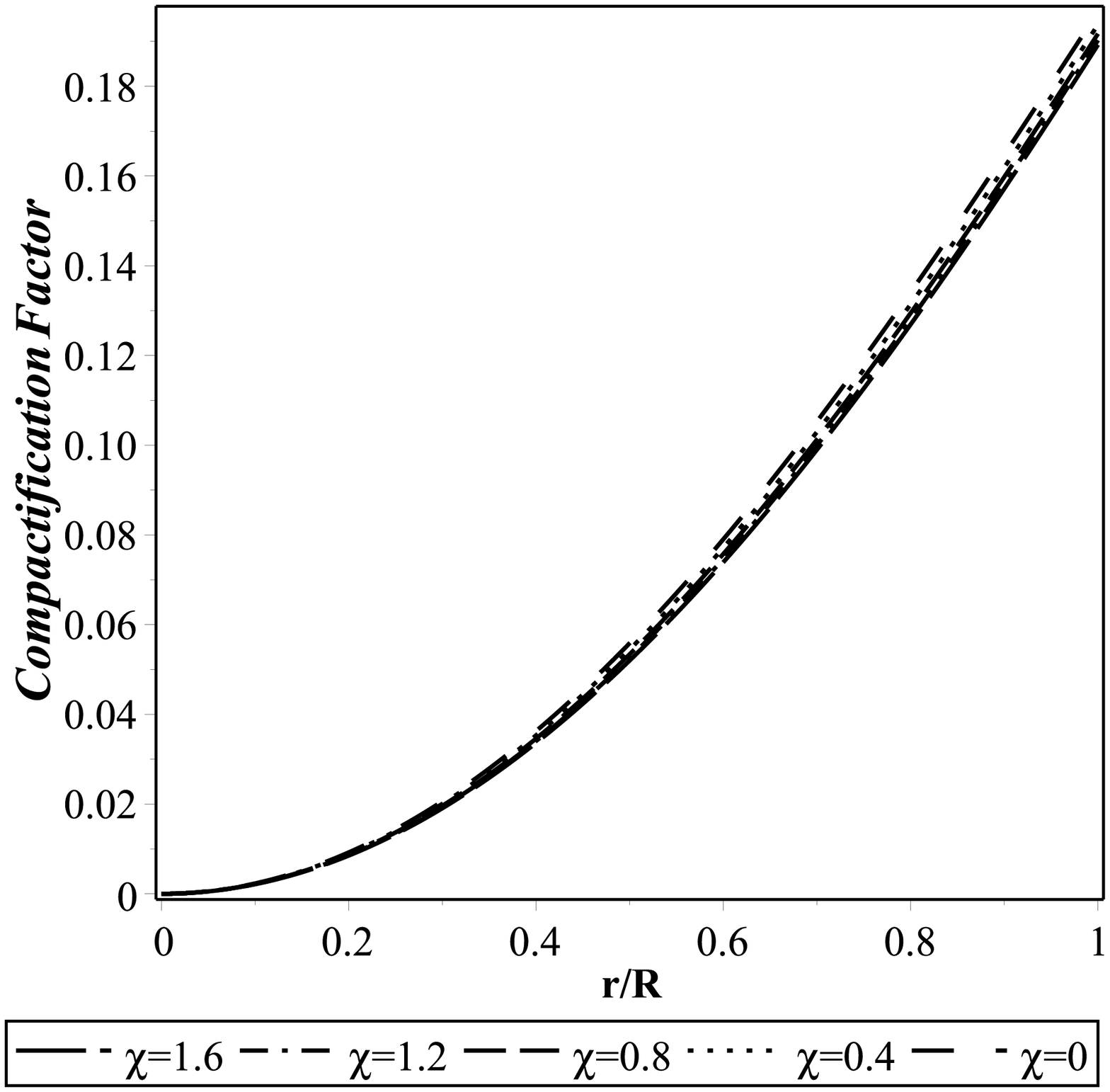}
\includegraphics[scale=.3]{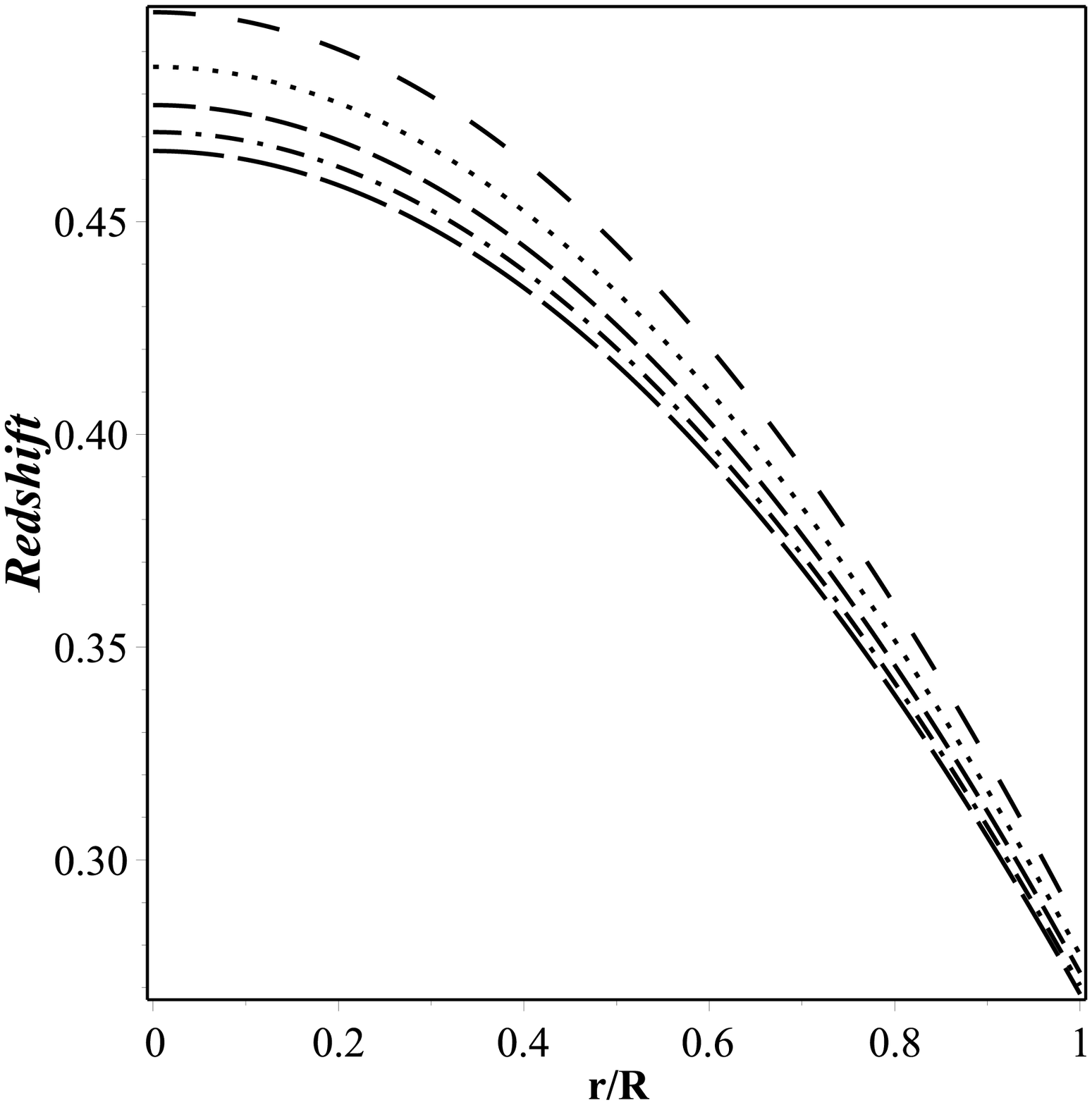}
\caption{Variation of the (i) compactification factor (upper panel) and (ii) redshift (lower panel) as a function of the radial coordinate $r/R$ for the strange star $LMC~X-4$.} \label{Fig10}
\end{figure}


We have featured the variation of the compactification factor and the redshift function with respect to the radial coordinate $r/R$ in Fig.~\ref{Fig10}.

\section{Discussion and Conclusion}\label{sec5}
Present article serves our motivation to explore the possibility of existence of the anisotropic ultra dense strange quark stars in the framework of the $f\left(R,\mathcal{T}\right)$ theory of gravity. To this end, following Harko et al.~\cite{Harko2011} we have considered simplified linear form of the arbitrary function $f\left(R,\mathcal{T}\right)$ given as $f(R,\mathcal{T})=R+2\chi\mathcal{T}$. In Eq.~(\ref{1.5}) we present the field equation due to the modified EH action in $f\left(R,\mathcal{T}\right)$ gravity. Eq.~(\ref{1.5}) clearly indicates that our system is not made of only the SQM but also a second kind of unknown matter is produced as a coupling effect of the matter and geometry. In this context, one may consult the article by Chakraborty~\cite{SC2013}, where he studied the nature and origin of this second kind of matter distribution which has been produced due to the effect of $f\left(R,\mathcal{T}\right)$ gravity. 

Now considering the stress-energy tensor due to the effective matter distribution as $T_{\mu\nu,eff}$ we find the standard form of the energy conservation equation in Eq.~(\ref{1.7}). To solve the Einstein field equations [(\ref{2.2})-(\ref{2.4})] we have considered that the SQM matter distribution is governed by the simplified MIT Bag EOS (\ref{2.10}) and assumed a relation between $p_t$ and $\rho$, as given in Eq.~(\ref{2.11a}). Throughout the study we have been considering $B=83~MeV/{{fm}^3}$~\cite{Rahaman2014}, $c_1=0.2$ and the values of $\chi$ as $0$,~$0.4$,~$0.8$,~$1.2$~and~$1.6$. We have illustrated the results graphically for $LMC~X-4$ as the representative of the strange quark stars.

We have presented features of the metric potentials, viz., $e^{\nu}$ (in the upper panel) and $e^{\lambda}$ (in the lower panel) in Fig.~\ref{Fig1}, which indicate that our stellar model is free from the geometrical singularity. In Fig.~\ref{Fig2} we have shown the variation of the energy density $({\rho}_{eff})$, radial pressure $(p_{effr})$ and tangential pressure $(p_{efft})$ in the upper, middle and lower panel, respectively. We find that ${\rho}_{eff}$,~$p_{effr}$ and $p_{efft}$ are maximum at the center and decrease monotonically inside the spherical system to achieve the minimum value on the surface. The anisotropy of the system is featured in Fig.~\ref{Fig3}, which shows that anisotropy is minimum, i.e., zero at the center and maximum on the surface in the present $f\left(R,\mathcal{T}\right)$ gravity model as the prediction made by Deb et al.~\cite{Deb2017} in the case of GR that maximum anisotropy on the surface is the inherent property of the anisotropic strange stars.


\begin{table*}[htbp!]
  \centering
    \caption{Numerical values of physical parameters for the strange star $LMC~X-4$ having mass~$1.29 \pm 0.05~M_{\odot}$~\cite{dey2013} for different $\chi$ } \label{Table 1}
    \begin{ruledtabular}
        \begin{tabular}{cccccccccccccccccc}
    Values of $\chi$ & $\chi=0$ & $\chi=0.4$ & $\chi=0.8$ & $\chi=1.2$ & $\chi=1.6$ &\\ 
\hline 
Predicted Radius \textit{(Km)} & 9.678 & 9.821 & 9.926 & 10.001 & 10.054 &\\
 \
${\rho}_{effc}$ ($gm/{{cm}^3}$) & $7.985\times {{10}^{14}}$ & $7.535\times {{10}^{14}}$ & $7.213\times {{10}^{14}}$ & $6.986\times {{10}^{14}}$ & $6.820\times {{10}^{14}}$ &\\ 
\
${\rho}_{eff0}$ ($gm/{{cm}^3}$) & $5.927\times {{10}^{14}}$ & $5.743\times {{10}^{14}}$ & $5.619\times {{10}^{14}}$ & $5.538\times {{10}^{14}}$ & $5.488\times {{10}^{14}}$ &\\
\
${p}_{effc}$ ($dyne/{{cm}^2}$) & $6.166\times {{10}^{34}}$ & $5.961\times {{10}^{34}}$ & $5.822\times {{10}^{34}}$ & $5.755\times {{10}^{34}}$ & $5.724\times {{10}^{34}}$ &\\ 
\
${2M}/{R}$ & 0.393 & 0.387 & 0.383 & 0.381 & 0.379 &\\
\ 
$Z_s$ & 0.284 & 0.277 & 0.273 & 0.271 & 0.269 & 
  \end{tabular}
    \end{ruledtabular}
    \end{table*}



\begin{table*}[htbp!]  \centering
    \caption{Numerical values of physical parameters for the different strange stars for $\chi=0.8$ and $c_1=0.2$ } \label{Table 2}
    \begin{ruledtabular}
    \begin{tabular}{cccccccccccccccc}
    Strange  & Observed & Predicted  & ${{\rho}_{effc}}$ & ${{\rho}_{eff0}}$ & ${{p}_{effc}}$ & Surface & $\frac{2M}{R}$ \\ 
  Stars &  Mass ($M_{\odot}$) &  Radius $(Km)$  &  $(gm/{cm}^3)$  &  $(gm/{cm}^3)$ &  $(dyne/{cm}^2)$ &  Redshift & \\ 
\hline 
$Vela~X-1$ & $1.77 \pm 0.08$~\cite{dey2013} & $10.866 \pm 0.133$ & $7.965 \times {{10}^{14}}$ & $5.597 \times {{10}^{14}}$ & $8.646 \times {{10}^{34}}$ & 0.388 & 0.481 \\ 
\ 
$4U~1820-30$ & $1.58 \pm 0.06$~\cite{guver2010b} & $10.529 \pm 0.113$ & $7.642\times {{10}^{14}}$ & $5.606\times {{10}^{14}}$ & $7.433\times {{10}^{34}}$ & 0.340 & 0.443 \\ 
\ 
$Cen~X-3$ & $1.49 \pm 0.08$~\cite{dey2013} & $10.354 \pm 0.160$ & $7.502\times {{10}^{14}}$ & $5.611\times {{10}^{14}}$ & $6.907\times {{10}^{34}}$ & 0.319 & 0.425 \\ 
\
$LMC~X - 4$ & $1.29 \pm 0.05$~\cite{dey2013} & $9.926 \pm 0.115$ & $7.213\times {{10}^{14}}$ & $5.619\times {{10}^{14}}$ & $5.822\times {{10}^{34}}$ & 0.273 & 0.383 \\ 
\
$SMC~X - 1$ & $1.04 \pm 0.09$~\cite{dey2013} & $9.299 \pm 0.248$ & $6.895\times {{10}^{14}}$ & $5.628 \times {{10}^{14}}$ & $4.624 \times {{10}^{34}}$ & 0.222 & 0.330 
  \end{tabular}
  \end{ruledtabular}
  \end{table*}


To examine the physical acceptability of the proposed anisotropic stellar model in $f\left(R,\mathcal{T}\right)$ gravity, we have studied the energy conditions, mass-radius relation, stability of the stellar system, etc. Fig.~\ref{Fig4} features that our system is consistent with all the energy conditions. Further, in Fig.~\ref{Fig5} we have presented the total mass $M$ (normalized in $M_{\odot}$) versus the total radius $R$ relations for the chosen values of $\chi$ and $B=83~MeV/{{fm}^3}$. The solid circles in Fig.~\ref{Fig5} are representing the maximum mass points for the strange stars. The figure shows that as the value of $\chi$ increases the value of $M$ and $R$ also gradually increases and thus provides a proportionality relation. We find for $\chi=1.6$ the value of $M_{max}$ increases to $17.36\%$ and $R_{Mmax}$ increases to $12.2\%$ than its corresponding value in GR and becomes $M_{max}=3.464~{M_{\odot}}$ and $R_{Mmax}=11.774~km$, respectively. In Fig.~\ref{Fig6} we have presented variation of $M$ (in the upper panel) and $R$ (in the lower panel) with respect to the central density of the effective matter distribution, ${\rho}_{effc}$. Fig.~\ref{Fig6} shows that with increase of $\chi$ the value of density decreases. For example, due to $\chi=1.6$ the maximum mass point $M_{max}=3.464~{M_{\odot}}$ is achieved for ${{\rho}_{effc}}=7.739\,{{\rho}_{nuclear}}$, which is $16.823\%$ lower than its value in GR. So, as the value of $\chi$ increases the strange stars become massive and bigger and thus show a gradual derease in its density.

To show stability of the system in terms of the equilibrium of forces we have studied modified TOV equation in the framework of $f\left(R,\mathcal{T}\right)$ theory of gravity. The variation of all the forces are featured in Fig.~\ref{Fig7}, which confirms that our system is stable in terms of the equilibrium of forces. Fig.~\ref{Fig7} also features an interesting fact that an extra force $F_m$ is produced due to the coupling effect between the matter and geometry. We introduced this force, $F_m$, as the modified force. We find that $F_m$ is repulsive in the nature and acts along the outward direction in the stellar system. To examine stability we also studied the Herrera cracking concept~\cite{Herrera1992,Abreu2007} and presented variation of the difference in square of the sound speeds, $|{v^2_{st}}- {v^2_{sr}}|$ against the radial coordinate $r/R$ in Fig.~\ref{Fig8}. We found our system is consistent with the causality condition and the Herrera cracking concept. Further, in Fig.~\ref{Fig9} we presented variation of both the adiabatic indices ${\Gamma}_r$ and ${\Gamma}_t$ with respect to the radial coordinate $r/R$ and have concluded that as both the adiabatic indices ${\Gamma}_r$ and ${\Gamma}_t$ are greater than $4/3$ so our system is stable against the radial pulsation. We also presented the variation of the compactification factor and the redshift in the upper and the lower panel, respectively in Fig.~\ref{Fig10}.

In TABLE~\ref{Table 1} we have predicted different physical parameters for the observed values of the mass of $LMC\,X-4$ for $B=83~MeV/{{fm}^3}$~\cite{Rahaman2014}, $c_1=0.2$ and the chosen values of $\chi$ as $0$,~$0.4$,~$0.8$,~$1.2$~and~$1.6$. We find as the coupling parameter, $\chi$ increases the mass~$(M)$ and the radius~$(R)$ of the star also increases gradually. However, TABLE~\ref{Table 1} shows that with the increasing value of $\chi$ the central~$({\rho}_{effc})$ and surface density~${\rho}_{eff0}$, central pressure~${p}_{effc}$, surface redshift~$(Z_s)$ and the value of $2M/R$ decreases gradually. Again, in TABLE~\ref{Table 2} we have presented above-mentioned physical parameters for different strange star candidates due to $\chi=0.8$. The high surface redshift values~$(0.388-0.222)$ and surface density values $(7.965 \times {{10}^{14}}-6.895 \times {{10}^{14}}~gm/{cm}^3 )$ as presented in TABLE~\ref{Table 2} clearly indicate that the stellar candidates are ultra-dense strange stars~\cite{Ruderman1972,Glendenning1997,Herjog2011}. It is also clear from both TABLES~\ref{Table 1} and~\ref{Table 2} that for all the values of $\chi$ our system is consistent with the Buchdahl condition~\cite{Buchdahl1959}, which demands a stringent condition $2M/R<8/9$. Now, as the compact stellar systems become gradually massive with the increment of $\chi$, hence our study reveals that the modified $f\left(R,\mathcal{T}\right)$ theory of gravity is a suitable theory to explain massive stellar systems like recent magnetars, massive pulsars and super-Chandrasekhar stars, which can not be explained in the framework of GR.

Again, by introducing $f\left(R,\mathcal{T}\right)=R+2\chi \mathcal{T}$ to consider the simplest minimal matter-geometry coupling, we have presented a similar and interesting result as presented by Astashenok et al.~\cite{Astashenok2015}. The authors~\cite{Astashenok2015} in their study presented a nonperturbative model of strange stars in $f\left(R\right)=R+\alpha {R}^2$ theory of gravity, where $\alpha$ is a constant. They showed that as the value of the constant parameter $\alpha$ increases mass of the strange star candidates increases gradually. In our present study we have also obtained the similar result for the increasing values of $\chi$, i.e., as the value of $\chi$ increases the stellar system becomes more massive gradually. It is interesting to note that the extra gravitational mass was arising in the case of $f\left(R\right)$~\cite{Astashenok2015} theory of gravity due to the extra geometrical term $\alpha {R}^2$, whereas in our study the same is obtained due to the extra material term $2\chi \mathcal{T}$. Hence, it is difficult to distinguish the effect of both the extra geometrical term and the material term on the ultra dense stellar configuration. 

However, in the case of $f\left(R\right)$ gravity Astashenok et al.~\cite{Astashenok2015} have obtained the maximum mass points due to the different values of $\alpha~(>0)$ for the higher values of central densities as compared to GR. On the contrary, in our study we find that for the different values of $\chi~(>0)$ the maximum mass points are achieved for the lower values of central densities as compared to GR. Also, Astashenok et al.~\cite{Astashenok2015} showed that with the increasing values of $\alpha$ from $\alpha=0$ in  $f\left(R\right)$ gravity  values of the surface redshift increases gradually, whereas we find with the increasing values of $\chi$ from $\chi=0$ in  $f\left(R, \mathcal{T}\right)$ theory of gravity the surface redshift decreases gradually. Thus by studying the central density and the surface redshift one may easily distinguish the effects and predictions of $f\left(R\right)$ and $f\left(R, \mathcal{T}\right)$ theory of gravities.

We can easily discriminate modified $f\left(R, \mathcal{T}\right)$ gravity from GR by noting at the surface redshift which has an inverse relationship with the parameter $\chi$. We also find that a stellar system becomes more massive in modified $f\left(R, \mathcal{T}\right)$ gravity compared to GR. It is worth mentioning that likewise GR in the present extended gravity theory too MIT bag model takes a  suitable role to discuss strange star candidates.

As a final comment, in this paper we have successfully presented a stable and physically acceptable anisotropic stellar model, which is suitable to study ultra-dense strange stars in the framework of $f\left(R,\mathcal{T}\right)$ theory of gravity.

\section*{ACKNOWLEDGMENTS}
SR and FR are thankful to the Inter-University Centre for Astronomy and Astrophysics (IUCAA), Pune, India for providing Visiting Associateship under which a part of this work was carried out. SR is also thankful to the authority of The Institute of Mathematical Sciences, Chennai, India for providing all types of working facility and hospitality under the Associateship scheme. FR is also grateful to DST-SERB (EMR/2016/000193), Govt. of India for providing financial support. A part of this work was completed while DD was visiting IUCAA and the author gratefully acknowledges the warm hospitality and facilities at the library there. We all are thankful to the anonymous referee for the pertinent comments which has helped us to upgrade the manuscript substantially.

\vspace{1.0cm}

\section*{APPENDIX: Expressions of constants}\label{appn}
The expressions of the constants used in Eqs.~(\ref{3.1})-(\ref{3.5}) are given as
\begin{eqnarray}\label{appn1}
&\qquad\hspace{-2cm} \lambda_{{1}}= \left( \frac{3}{4}c_{{1}}+\frac{3}{4} \right) \chi+\pi, \\ \label{appn2}
&\qquad\hspace{-2cm}  \lambda_{{2}}= \left( -\frac{1}{4}c_{{1}}+\frac{1}{12}\right) \chi+\pi, \\ \label{appn3}
&\qquad\hspace{-1cm}  \lambda_{{3}}=-384\, \left[  \left( \frac{5}{8}B-\frac{1}{4}\rho_{{c}} \right) 
 \chi+\pi \, \left( B-\frac{1}{4}\rho_{{c}} \right)  \right] \lambda_{{2}}, \\ \label{appn4}
&\qquad\hspace{-1cm} \nu_{{1}}= \left( \frac{3}{8}c_{{1}}-\frac{1}{8} \right) {\chi}^{2}+\frac{9}{4}\left( c_{{1}}+\frac{1}{9}\right) \pi \,\chi+{\pi }^{2}, \\ \label{appn5}
&\qquad\hspace{-1cm} \nu_{{2}}= \left( \pi +\frac{\chi}{2}\right)  \left( \pi +\frac{\chi}{4}\right), \\ \label{appn5a}
&\qquad\hspace{-1cm} \nu_{{3}}=589824\, \left( \pi +\frac{\chi}{4} \right) {R}^{2} \Bigg[ B{\pi }^{3}{R}^{3}+ \Big\lbrace \frac{1}{8}B{R}^{3} \left( c_{{1}}+{\frac {26}{3}} \right) \chi \nonumber\\
&\qquad\hspace{-1cm} -{\frac {5M}{32}} \Big\rbrace {\pi }^{2}-\frac {3\chi\pi }{32} \Big\lbrace {R}^{3} \left( {c_{{1}}}^{2}+\frac{1}{3}c_{{1}}-{\frac {32}{9}} \right) B\chi-\frac{5}{4}\left( c_{{1}}
-\frac{5}{3}\right) M \Big\rbrace\nonumber\\
&\qquad\hspace{-1cm}-\frac {3\,{\chi}^{2} \left( c_{{1}}-\frac{1}{3}\right) }{64} \Big\lbrace B{R}^{3} \left( c_{{1}}+\frac{4}{3}\right) \chi-{\frac {15\,M \left( c_{{1}}+1 \right) }{4}} \Big\rbrace  \Bigg], \\ \label{appn6}
&\qquad\hspace{-1cm} \nu_{{4}}={R}^{2}\Big[{\nu_{{2}}}^{2}{\lambda_{{2}}}^{2}{B}^{2}{R}^{6}+\frac{1}{4}\nu_{{1}}\nu_{{2
}}\lambda_{{2}}B{R}^{4}-\frac{5}{8}\nu_{{2}}\lambda_{{1}}\pi \,M\lambda_{{2}
}B{R}^{3}\nonumber\\
&\qquad\hspace{-1cm}-{\frac {3\,\nu_{{1}}M \left( \pi +\chi \right) \lambda_{{2}}
R}{64}}+{\frac {25\,{\lambda_{{1}}}^{2}{\pi }^{2}{M}^{2}}{256}}
\Big]^{\frac{1}{2}}, \\ \label{appn7}
&\qquad\hspace{-1cm} \nu_{{5}}=24{R}^{5} \Big[\frac{9}{4}\chi\left(\pi +\frac{\chi}{6}\right) c_{{1}}
+ \left( \pi +\frac{\chi}{2}\right)\left( \pi -\frac{\chi}{4}\right)\Big], \\ \label{appn8}
&\qquad\hspace{-1cm} p_{{1}}=16\,B\pi +10\,B\chi-4\,\pi \,\rho_{{c}}-4\,\chi\,\rho_{{c}}, \\ \label{appn9}
&\qquad\hspace{-1cm} {{\rho}_c}=-\Big[\Big\lbrace-48\,B\pi \,{R}^{3}\chi\,c_{{1}}+192\,B{\pi }^{2}{R}^{3}+176\,B\pi \,{R}^{3}\chi\nonumber\\
&\qquad\hspace{-1cm}+40\,B{R}^{3}{\chi}^{2}-45\,M\chi\,c_{{1}}-60\,M\pi -45\,M\chi\Big\rbrace\Big/\nonumber\\
&\qquad\hspace{-1cm}\Big\lbrace 4\,{R}^{3} \left( 18\,\pi \,\chi\,c_{{1}}+3\,{\chi}^{2}c_{{1}}+8\,{\pi }^{2}+2\,\pi \,\chi-{\chi}^{2} \right)\Big\rbrace\Big],\\ \label{appn10}
&\qquad\hspace{-1cm} \rho_{{0}}={\frac {3\,\chi\,c_{{1}}\rho_{{c}}+16\,B\pi +10\,B\chi-\chi
\,\rho_{{c}}}{3\,\chi\,c_{{1}}+4\,\pi +3\,\chi}}.
\end{eqnarray}

\begin{thebibliography}{99}

\bibitem{Riess1998} A.G. Riess et al., Astron. J. \textbf{116}, 1009 (1998).

\bibitem{Perlmutter1999} S. Perlmutter et al., Astrophys. J. \textbf{517}, 565  (1999).

\bibitem{Bernardis2000} P. de Bernardis et al, Nature \textbf{404}, 955 (2000).

\bibitem{Perlmutter2003} R.A. Knop et al., Astrophys. J. \textbf{598}, 102 (2003).

\bibitem{Caldwell2002} R.R. Caldwell, Phys. Lett. B \textbf{545}, 23 (2002).

\bibitem{Nojiri2003} S. Nojiri and S.D. Odinstov, Phys. Lett. B \textbf{562}, 147 (2003).

\bibitem{Odinstov2003} S. Nojiri and S.D. Odinstov, Phys. Lett. B \textbf{ 565}, 1 (2003).

\bibitem{Padmanabhan2002} T. Padmanabhan and T.R. Choudhury, Phys. Rev. D \textbf{66}, 081301 (2002).

\bibitem{Kamenshchik2001} A. Kamenshchik, U. Moschella, and V. Pasquier, Phys. Lett. B \textbf{511}, 265 (2001).

\bibitem{Bento2002} M.C. Bento, O. Bertolami, and A.A. Sen, Phys. Rev. D \textbf{66}, 043507 (2002).

\bibitem{Nojiri2011} S. Nojiri and S.D. Odintsov, Phys. Rep. \textbf{505}, 59 (2011).

\bibitem{Nojiri2003a} S. Nojiri and S.D. Odintsov, Phys. Rev. D \textbf{68}, 123512 (2003).

\bibitem{Carroll2004} S.M. Carroll, V. Duvvuri, M. Trodden, and M.S. Turner, Phys. Rev. D \textbf{70}, 043528 (2004).

\bibitem{Allemandi2005} G. Allemandi, A. Borowiec, M. Francaviglia, and S.D. Odintsov, Phys. Rev. D \textbf{72}, 063505 (2005).

\bibitem{Nojiri2007} S. Nojiri and S.D. Odintsov, Int. J. Geom. Meth. Mod. Phys. \textbf{04}, 115 (2007).

\bibitem{Bertolami2007} O. Bertolami, C.G. Bohmer, T. Harko, and F.S.N. Lobo, Phys. Rev. D \textbf{75}, 104016 (2007).

\bibitem{Bamba2010a} K. Bamba, C.Q. Geng, S. Nojiri, and S.D. Odintsov, EPL \textbf{89}, 50003 (2010).

\bibitem{Bamba2010b} K. Bamba, S.D. Odintsov, L. Sebastiani, and S. Zerbini, Eur. Phys. J. C \textbf{67}, 295 (2010).

\bibitem{Rodrigues2014} M.E. Rodrigues, M.J.S. Houndjo, D. Mommeni, and R. Myrzakulov, Can. J. Phys. \textbf{92}, 173 (2014).

\bibitem{Bengocheu2009} G.R. Bengochea and R. Ferraro, Phys. Rev. D \textbf{79}, 124019 (2009).

\bibitem{Linder2010} E.V. Linder, Phys. Rev. D \textbf{81}, 127301 (2010).

\bibitem{Capozziello2002} S. Capozziello, Int. J. Mod. Phys. D, \textbf{11}, 483 (2002).

\bibitem{Capozziello2011} S. Capozziello and M. DeLaurentis, Phys. Rep. \textbf{509}, 167 (2011).

\bibitem{Astashenok2013} A.V. Astashenok, S. Capozziello, and S.D. Odintsov, J. Cosmol. Astropart. Phys. 12 (2013) 040.

\bibitem{Capozziello2016} S. Capozziello, M. DeLaurentis, R. Farinelli, and S.D. Odintsov, Phys. Rev. D \textbf{93}, 023501 (2016).

\bibitem{Astashenok2015} A.V. Astashenoka, S. Capozziello, and S.D. Odintsov, Phys. Lett. B  \textbf{742}, 160 (2015).

\bibitem{Harko2011} T. Harko, F.S.N. Lobo, S. Nojiri, and S.D. Odintsov, Phys. Rev. D \textbf{84}, 024020 (2011).

\bibitem{Myrzakulov2012} R. Myrzakulov, Eur. Phys. J. C \textbf{72}, 2203 (2012).

\bibitem{Jamil2012} M. Jamil, D. Momeni, and R. Myrzakulov, Chin. Phys. Lett. \textbf{29}, 109801 (2012).

\bibitem{Shabani2013} H. Shabani and M. Farhoudi, Phys. Rev. D \textbf{88}, 044048 (2013).

\bibitem{Shabani2014} H. Shabani and M. Farhoudi, Phys. Rev. D \textbf{90}, 044031 (2014).

\bibitem{Moraes2015d} P.H.R.S. Moraes, Eur. Phys. J. C \textbf{75}, 168 (2015).

\bibitem{Momeni2015} D. Momeni, R. Myrzakulov, and E. G{\"u}dekli, Int. J. Geom. Meth. Mod. Phys. \textbf{12}, 1550101 (2015).

\bibitem{Zaregonbadi2016} R. Zaregonbadi and M. Farhoudi, Gen. Rel. Grav. \textbf{48}, 142 (2016).

\bibitem{Shabani2017a} H. Shabani, Int. J. Mod. Phys. D \textbf{26}, 1750120 (2017).
        
\bibitem{Shabani2017b} H. Shabani and A.H. Ziaie, Eur. Phys. J. C \textbf{77}, 31 (2017).

\bibitem{sharif2014} M. Sharif and Z. Yousaf, Astrophys. Space Sci. {\bf 354}, 471 (2014).

\bibitem{noureen2015} I. Noureen and M. Zubair, Astrophys. Space Sci. {\bf 356}, 103 (2015).

\bibitem{noureen2015b} I. Noureen and M. Zubair, Eur. Phys. J. C {\bf 75}, 62 (2015).

\bibitem{noureen2015c} I. Noureen, M. Zubair, A.A. Bhatti, and G. Abbas, Eur. Phys. J. C {\bf 75}, 323 (2015).

\bibitem{zubair2015a} M. Zubair and I. Noureen, Eur. Phys. J. C \textbf{75}, 265 (2015).

\bibitem{zubair2015b} M. Zubair, G. Abbas, and I. Noureen, Astrophys. Space Sci. \textbf{361}, 8 (2016).

\bibitem{Ahmed2015} A. Alhamzawi and R. Alhamzawi, Int. J. Mod. Phys. D {\bf 25}, 1650020 (2016).

\bibitem{Psaltis2008} D. Psaltis, Living Rev. Relativ. \textbf{11}, 9 (2008).

\bibitem{Capozziello2011a} S. Capozziello, M. DeLaurentis, S.D. Odintsov, and A. Stabile, Phys. Rev. D \textbf{83}, 064004 (2011).

\bibitem{Capozziello2012} S. Capozziello, M. DeLaurentis, I. DeMartino, M. Formisano, and S.D. Odintsov, Phys. Rev. D \textbf{85}, 044022 (2012).

\bibitem{Briscese2007} F. Briscese, E. Elizalde, S. Nojiri, and S.D. Odintsov, Phys. Lett. B \textbf{646}, 105 (2007).

\bibitem{Abdalla2005} M.C.B. Abdalla, S. Nojiri, and S.D. Odintsov, Class. Quantum Gravity \textbf{22}, L35 (2005).

\bibitem{Bamba2008} K. Bamba, S. Nojiri, and S.D. Odintsov, J. Cosmol. Astropart. Phys. 10 (2008) 045.

\bibitem{Kobayashi2008} T. Kobayashi and K.I. Maeda, Phys. Rev. D \textbf{78}, 064019 (2008).

\bibitem{Babichev2010} E. Babichev and D. Langlois, Phys. Rev. D \textbf{81}, 124051 (2010).

\bibitem{Nojiri2009} S. Nojiri and S.D. Odintsov, Phys. Lett. B \textbf{676}, 94 (2009).

\bibitem{Bamba2011} K. Bamba, S. Nojiri, and S.D. Odintsov, Phys. Lett. B \textbf{698}, 451 (2011).

\bibitem{Khoury2004a} J. Khoury and A. Weltman, Phys. Rev. D \textbf{69}, 044026 (2004).

\bibitem{Khoury2004b} J. Khoury and A. Weltman, Phys. Rev. Lett. \textbf{93}, 171104 (2004).

\bibitem{Upadhye2009} A. Upadhye and W. Hu, Phys. Rev. D \textbf{80}, 064002 (2009).

\bibitem{Moraes2015} P.H.R.S. Moraes, J.D.V. Arba{\~n}il, and M. Malheiro, J. Cosmol. Astropart. Phys. 06 (2016) 005.

\bibitem{Amit2016} A. Das, F. Rahaman, B.K. Guha, and S. Ray, Eur. Phys. J. C \textbf{76}, 654 (2016).

\bibitem{Amit2017} A. Das, S. Ghosh, B.K. Guha, S. Das, F. Rahaman, and S. Ray, Phys. Rev. D \textbf{95}, 124011 (2017). 

\bibitem{Harko2014} T. Harko, Phys. Rev. D \textbf{90}, 044067 (2014).

\bibitem{SC2013} S. Chakraborty, Gen. Relativ. Gravit. \textbf{45}, 2039 (2013).

\bibitem{Ivanov2002} B.V. Ivanov, Phys. Rev. D \textbf{65}, 104011 (2002).

\bibitem{SM2003} F.E. Schunck and E.W. Mielke, Class. Quantum Gravit. \textbf{20}, R301 (2003).

\bibitem{MH2003} M.K. Mak and T. Harko, Proc. R. Soc. A \textbf{459}, 393 (2003).

\bibitem{Usov2004} V.V. Usov, Phys. Rev. D \textbf{70}, 067301 (2004).

\bibitem{Varela2010} V. Varela, F. Rahaman, S. Ray, K. Chakraborty, and M. Kalam, Phys. Rev. D \textbf{82}, 044052 (2010).

\bibitem{Rahaman2010} F. Rahaman, M. Jamil, A. Ghosh, and K. Chakraborty, Mod. Phys. Let. A \textbf{25}, 835 (2010).

\bibitem{Rahaman2011} F. Rahaman, P.K.F. Kuhfittig, M. Kalam, A.A. Usmani, and S. Ray, Class. Quantum Gravit. \textbf{28}, 155021 (2011).

\bibitem{Rahaman2012} F. Rahaman, R. Maulick, A.K. Yadav, S. Ray, and R. Sharma, Gen. Relativ. Gravit. \textbf{44}, 107 (2012).

\bibitem{Kalam2012} M. Kalam, F. Rahaman, S. Ray, Sk.M. Hossein, I. Karar, and J. Naskar, Eur. Phys. J. C \textbf{72}, 2248 (2012).

\bibitem{Deb2016} D. Deb, S. R. Chowdhury, B.K. Guha, and S. Ray, arXiv:1611.02253 [gr-qc].
	
\bibitem{Shee2016} D. Shee, F. Rahaman, B.K. Guha, and S. Ray, Astrophys. Space Sci. \textbf{361}, 167 (2016).

\bibitem{Maurya2016} S.K. Maurya, Y.K. Gupta, S. Ray, and D. Deb, Eur. Phys. J. C \textbf{76}, 693 (2016).

\bibitem{Maurya2017} S.K. Maurya, D. Deb, S. Ray, and P.K.F. Kuhfittig, eprint arXiv:1703.08436.

\bibitem{singh2015} V. Singh and C.P. Singh, Astrophys. Space Sci. {\bf 356}, 153 (2015).

\bibitem{moraes2014b} P.H.R.S. Moraes, Astrophys. Space Sci. \textbf{352}, 273 (2014).

\bibitem{moraes2015a} P.H.R.S. Moraes, Eur. Phys. J. C \textbf{75}, 168 (2015).

\bibitem{moraes2015b} P.H.R.S. Moraes, Int. J. Theor. Phys. \textbf{55}, 1307 (2016).

\bibitem{moraes2017} P.H.R.S. Moraes, R.A.C. Correa, and R.V. Lobato, J. Cosmol. Astropart. Phys. 07 (2017) 029.

\bibitem{singh2014} C.P. Singh and P. Kumar, Eur. Phys. J. C \textbf{74}, 3070 (2014).

\bibitem{baffou2015} E.H. Baffou, A.V. Kpadonou, M.E. Rodrigues, M.J.S. Houndjo, and J. Tossa, Astrophys. Space Sci. {\bf 356}, 173 (2015).

\bibitem{shabani2013} H. Shabani and M. Farhoudi, Phys. Rev. D \textbf{88}, 044048 (2013).

\bibitem{shabani2014} H. Shabani and M. Farhoudi, Phys. Rev. D \textbf{90}, 044031 (2014).

\bibitem{sharif2014b} M. Sharif and M. Zubair, Astrophys. Space Sci. {\bf 349}, 457 (2014).

\bibitem{reddy2013b} D.R.K. Reddy and R. Shantikumar, Astrophys. Space Sci. {\bf 344}, 253 (2013).

\bibitem{kumar2015} P. Kumar and C.P. Singh, Astrophys. Space Sci. {\bf 357}, 120 (2015).

\bibitem{shamir2015} M.F. Shamir, Eur. Phys. J. C {\bf 75}, 354 (2015).

\bibitem{Fayaz2016} V. Fayaz, H. Hossienkhani, Z. Zarei, and N. Azimi, Eur. Phys. J. Plus {\bf 131}, 22 (2016).

\bibitem{Chodos1974} A. Chodos, R.L. Jaffe, K. Johnson, C.B. Thorn, and V.F. Weisskopf, Phys. Rev. D \textbf{9}, 3471 (1974).

\bibitem{Rahaman2014} F. Rahaman, K. Chakraborty, P.K.F. Kuhfittig, G.C. Shit, and M. Rahman, Eur. Phys. J. C \textbf{74}, 3126 (2014).

\bibitem{1} M. Brilenkov, M. Eingorn, L. Jenkovszky, and A. Zhuk, J. Cosmol. Astropart. Phys. 08 (2013) 002.

\bibitem{2} N.R. Panda, K.K. Mohanta, and P.K. Sahu, J. Physics: Conference Series \textbf{599}, 012036 (2015).

\bibitem{3} A.A. Isayev, Phys. Rev. C \textbf{91}, 015208 (2015).

\bibitem{4} S.D. Maharaj, J.M. Sunzu, and S. Ray, Eur. Phys. J. Plus. \textbf{129}, 3 (2014).

\bibitem{5} L. Paulucci and J.E. Horvath, Phys. Lett. B \textbf{733}, 164 (2014).

\bibitem{6}  G. Abbas, S. Qaisar, and A. Jawad, Astrophys. Space Sci. \textbf{359}, 57 (2015).

\bibitem{7} J.D.V. Arba{\~n}il and M. Malheiro, J. Cosmol. Astropart. Phys. 11 (2016) 012.

\bibitem{8} G. Lugones and J.D.V. Arba{\~n}il, Phys. Rev. D \textbf{95}, 064022 (2017).

\bibitem{Harko2002} M.K. Mak and T. Harko, Chin. J. Astron. Astrophys. \textbf{2}, 248 (2002).

\bibitem{Moraes2017} P.H.R.S. Moraes, R.A.C. Correa, and R.V. Lobato, J. Cosmol. Astropart. Phys. 07 (2017) 029.

\bibitem{Deb2017} D. Deb, S. Roy Chowdhury, S. Ray, F. Rahaman, and B.K. Guha, Ann. Phys. \textbf{387}, 239 (2017).

\bibitem{dey2013} T. Gangopadhyay, S. Ray, X.D. Li, J. Dey, and M. Dey, Mon. Not. R. Astron. Soc. {\bf 431}, 3216 (2013).

\bibitem{guver2010b} T. G{\"u}ver, F. {\"O}zel, A. Cabrera-Lavers, and P. Wroblewski, Astrophys J. {\bf 712}, 964 (2010).

\bibitem{Herrera1992} L. Herrera, Phys. Lett. A \textbf{165}, 206 (1992).

\bibitem{Abreu2007} H. Abreu, H. Her{\'n}andez, and L.A. N{{\'u}}${\tilde{n}}$ez, Class. Quantum Gravit. \textbf{24}, 4631 (2007).

\bibitem{Chandrasekhar1964} S. Chandrasekhar, Astrophys J. \textbf{140}, 417 (1964).

\bibitem{Hillebrandt1976} W. Hillebrandt and K.O. Steinmetz, Astron. Astrophys. \textbf{53}, 283 (1976). 

\bibitem{Chan1994} R. Chan, L. Herrera, and N.O. Santos, Mon. Not. R. Astron. Soc. \textbf{267}, 637 (1994).

\bibitem{Herrera1997} L. Herrera and N.O. Santos, Phys. Rep. \textbf{286}, 53 (1997).

\bibitem{Horvat2010} D. Horvat, S. Iliji\'c, and A. Marunovi\'c, Class. Quantum Gravit. \textbf{28}, 025009 (2011).

\bibitem{Doneva2012} D.D. Doneva and S.S. Yazadjiev, Phys. Rev. D \textbf{85}, 124023 (2012).

\bibitem{Silva2015} H.O. Silva, C.F.B. Macedo, E. Berti, and L.C.B. Crispino, Class. Quantum Gravit. \textbf{32}, 145008 (2015).

\bibitem{Heintzmann1975} H. Heintzmann and W. Hillebrandt, Astron. Astrophys. \textbf{38}, 51 (1975).

\bibitem{Ruderman1972} R. Ruderman, Rev. Astron. Astrophys. \textbf{10}, 427 (1972).

\bibitem{Glendenning1997} N.K. Glendenning, Compact Stars: Nuclear Physics, Particle Physics and General Relativity (Springer, New York, 1997).

\bibitem{Herjog2011} M. Herzog and F.K.R{\"o}pke, Phys. Rev. D \textbf{84}, 083002 (2011).

\bibitem{Buchdahl1959} H.A. Buchdahl, Phys. Rev. D \textbf{116}, 1027 (1959).

\end{thebibliography}
\end{document}